\documentclass[prb,showpacs]{revtex4}

\usepackage{graphicx}
\usepackage{epsfig}
\usepackage{bm}
\usepackage{latexsym}
\newcommand{\ket}[1]{| {#1} \rangle}
\newcommand{\vev}[1]{\mbox{$\left\langle #1 \right\rangle$}}

\makeatletter
 \def\varddots{\mathinner{\mkern1mu
     \raise\p@\hbox{.}\mkern2mu\raise4\p@\hbox{.}\mkern2mu
     \raise7\p@\vbox{\kern7\p@\hbox{.}}\mkern1mu}}
\makeatother

\makeatletter
 
  \@addtoreset{equation}{section}
\makeatother

\begin{document}

\title{Critical Spectra and Wavefunctions of a One-dimensional 
Quasiperiodic System }

\author{ Kazusumi Ino$^1$ and Mahito Kohmoto$^2$ }

\affiliation{
$^1$Department of Basic Science, University of Tokyo, 
Komaba 3-8-1, Meguro-ku, Tokyo, 153-8902, Japan\\
$^2$Institute of Solid State Physics, University of Tokyo,
Kashiwanoha 5-1-5, Kashiwa-shi, Chiba, 277-8581, Japan}

\begin{abstract}

We numerically study a one dimensional quasiperiodic system  
 obtained  from two dimensional electrons on the  
 triangular lattice in a uniform magnetic field aided by the multifractal method.
The phase diagram  
consists of three phases: two metallic phases and 
one insulating phase separated by critical lines 
with one bicritical point. Novel transitions between the two metallic phases exist.
We examine the spectra and the wavefunctions along  the critical lines. 
Several types of level statistics are obtained.  
Distributions of the band widths $P_B(w)$ near the origin (in the tail)
have a form  
$P_B(w) \sim w^{ \beta}$ ($P_B(w) \sim e^{ -\gamma w }$) 
($\beta , \gamma > 0 $),
while at the bicritical point $P_B(w) \sim w^{-\beta'}$ ($\beta'>0$). 
Also distributions of the level spacings follow 
an inverse power law $P_G(s) \sim s^{- \delta}$ ($\delta > 0$ ).
For the wavefunctions at the centers of spectra,  scaling exponents and their 
distribution in terms of  the $\alpha$-$f(\alpha)$-curve are obtained.  
The results 
in the vicinity of critical points are consistent with 
the phase diagram. 
\end{abstract}

\pacs{71.30.+h,  71.23.Ft, 05.45.Mt}

\maketitle

\section{Introduction}

A peculiar problem of 
two-dimensional  electrons  in  a periodic potential with 
a perpendicular magnetic field  
has been attracted much attentions 
since the Hofstadter butterfly \cite{hof}. 
It appears as the spectrum of the underlying one-dimensional 
system called the Harper model \cite{harper} which is deduced 
from  electrons on the square lattice in a uniform magnetic field .  It is also essential in physics of 
the integer quantum Hall effect \cite{tknn,kohmoto-chern}.

Some of the metal-insulator transitions  
in one-dimensional quasiperiodic systems have been characterized
by multifractal structures of band widths and wavefunctions.
See \cite{hiramoto}.
A one-dimensional tight-binding model is    
\begin{equation}
t_{i+1}\psi_{i+1}+t_{i-1}\psi_{i-1}+\epsilon_i \psi_i =E\psi_i,
\label{2}
\end{equation} 
where $\psi_i $ denotes  the value of 
the wavefunction at the $i$-th site, 
$t_i$ and $\epsilon_i$  are  the hopping matrix element 
and  the site energy at the $i$-th site respectively, either or both of them can   
be taken to be quasiperiodic. 
The Harper model is the case where
$t_{i}=1$ and $\epsilon_i=\lambda \cos(2\pi\sigma i+\theta)$. 
When $\sigma$ is an irrational number, it is quasiperiodic.  
All the eigenstates 
 are extended for $\lambda <2$ 
and are localized for $\lambda >2$ with the metal-insulator 
transitions at $\lambda=2$ \cite{aubry,kohmoto-prl}.  The spectrum has a rich  structure 
 (the Hofstadter butterfly). 
The spectrum as well as the eigenstates becomes multifractal
 %
%
\cite{hiramoto-kohmoto}. 
The total measure of the bands at the critical point $\lambda =2$
is zero with a fractal dimension less than one \cite{thouless}. 
The scaling behaviors of the spectra have been extensively studied   
 \cite{hiramoto-kohmoto}
by the multifractal analysis \cite{halsey,kohmoto}. 
Especially the incommensurate limits of flux per plaquette,
such as the inverse of the golden mean $\sigma=\frac{-1+\sqrt{5}}{2}$
have been extensively studied.

Recently level statistics of some of the quasiperiodic systems 
have been investigated and turned out to have the
behaviors  characteristic of criticality  \cite{evangelou,takada}. 
The distributions of the normalized band widths
$P_B(w)dw$ have been confirmed that 
\begin{eqnarray}
P_B(w)\sim w^{\beta}
\quad (w\to 0),
\end{eqnarray}
and
\begin{eqnarray}
P_B(w)\sim e^{-\gamma w}
\quad (w \to \infty),
\end{eqnarray}
These laws have also been confirmed for  
a variant of the Harper model at criticality 
\cite{takada}.  A similar type of statistical law has 
been confirmed for the Fibonacci model \cite{naka}. 
Remarkably, the form of the distributions of band widths has 
a similar form as the distributions of the gaps 
fluctuations observed at the mobility edge of the random systems \cite{altshuler,shapiro}. 
Distribution of the energy gaps    $P_G(s)$ 
 was also examined. 
It diverges near the origin and 
follows an inverse power law \cite{machigei,takada,naka}
\begin{eqnarray}
P_G(s)\sim s^{-\delta} \quad (s\to 0),
\end{eqnarray}
 For example, 
 $\delta \sim 1.5$ for the critical Harper model($\lambda=2$).

One of the  aims of this paper is to investigate these quantities 
for the one dimensional quasiperiodic model obtained 
from two dimensional electrons on the triangular lattice 
in a uniform magnetic field.  This problem was
 studied by Claro and Wannier \cite{claro} ,
but systematic studies have not been achieved since then.
 Although the model of two-dimensional 
 electrons on the square lattice with next-nearest hopping, 
 which includes the case of triangular lattice as a special case,
 were studied previously \cite{han-thouless,hat-koh}, 
  statistical techniques such  as the multifractal analysis 
 which have been applied for other quasiperiodic systems 
 have not been applied to the triangular lattice model.  
The same model  also 
appears in  the theory of the junction of three wires of Luttinger liquid 
\cite{chamon}.  These situations motivate us to investigate 
various aspects of the quasiperiodic system obtained 
from the triangular lattice model.

The organization of this paper is as follows. In Sec.{\ref{section;model}}, 
we introduce two dimensional electrons on the triangular lattice 
in a uniform magnetic field and obtain the one-dimensional 
quasiperiodic system. We 
 investigate the Aubry and Andr\'{e} duality \cite{aubry} in this model.  
In Sec.{\ref{section;chara}}, we investigate the classical orbits 
of the model and discuss the phase diagram.  
In Sec.{\ref{section;level}},  the distributions of the band widths and 
 the gaps are investigated. 
In Sec.{\ref{section;multi}}, we give a brief review of 
the general formulation of the multifractal analysis and 
apply it to the spectra and the wavefunctions.  We confirm 
the phase diagram conjectured in Sec.\ref{section;model}. 
Sec.{\ref{section;conclusions}}  is the  conclusion.

\section{\label{section;model}Electrons on the triangular lattice 
in a uniform magnetic field}
\subsection{Hamiltonian in real space}
\begin{figure}
  \begin{center}
    \epsfxsize=8cm
    \epsfbox{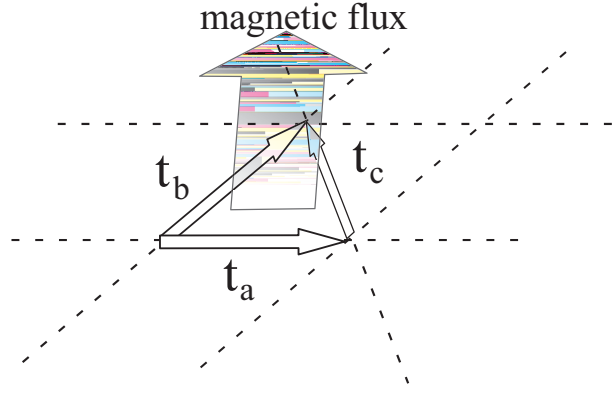}
    \caption{Schematic view of the triangular lattice.}. 
    \label{fig;tri}
  \end{center}
\end{figure}

We consider  tight-binding electrons on the triangular lattice 
in a magnetic field ({\bf  Fig.\ref{fig;tri}}). 
We take the lattice spacing to be $1$ for simplicity. 
The Hamiltonian is 
\begin{eqnarray}
H&=& -t_a\sum_{n,m}c^{\dagger}_{n+1,m}c_{n,m} \exp(iA_{n+1,m;n,m})
-t_a\sum_{n,m}c^{\dagger}_{n,m}c_{n+1,m} \exp(iA_{n,m;n+1,m})
 \nonumber \\
& & -t_b\sum_{n,m}  c^{\dagger}_{n,m+1}c_{n,m} \exp(iA_{n,m+1;n,m})
-t_b\sum_{n,m}  c^{\dagger}_{n,m}c_{n,m+1} \exp(iA_{n,m;n,m+1})
\nonumber \\
& &
-t_c\sum_{n,m} c^{\dagger}_{n,m+1}c_{n+1,m} \exp(iA_{n,m+1;n+1,m})
-t_c\sum_{n,m} c^{\dagger}_{n+1,m}c_{n,m+1} \exp(iA_{n+1,m;n,m+1})
\nonumber \\ 
&& \equiv H_a+H_b + H_c
\label{eq;model}
\end{eqnarray}
Here $t_a,t_b$ and $t_c$ are the hopping coefficients for each bond, 
and $c_{n,m}$($c^{\dagger}_{n,m}$) is the 
annihilation (creation) operator  at site $(n,m)$ : 
$\{c^{\dagger}_{n,m},c_{k,l} \}=\delta_{k,n}\delta_{lm}$. 
$A_{n,m;k,l}$, $k=n\pm1, l=m\pm1$
 is a gauge field on each bond. We impose  $A_{n,m;k,l} =-
A_{k,l;n,m}$ so that $H$ to be hermitian.  
A uniform magnetic field penetrates 
each triangle with a flux $\varphi=\frac{\phi}{2}$.  
We take the Landau gauge  
\begin{eqnarray}
A_{n+1,m;n,m} =0, A_{n,m+1;n,m} =2\pi \phi  \hspace{3mm}
{\rm and} \hspace{3mm}
 A_{n,m+1;n+1,m} =2\pi\phi (n+\frac{1}{2}).    
\label{t_a-gauge} 
\end{eqnarray}
thus $\sum_{{\rm triangle}} A_{n,m;k,l}=\frac{\phi}{2}$.
A state $\ket{\Psi}$ is written 
\begin{eqnarray}
\ket{\Psi} = \sum_{n,m} \Psi_{n,m}c^{\dagger}_{n,m}\ket{0}.  
\end{eqnarray} 
The Schr\"odinger equation $H\ket{\Psi}=E\ket{\Psi}$ is   
\begin{eqnarray}
-t_a(\Psi_{n-1,m}+\Psi_{n+1,m})-t_b(e^{2\pi i\phi n}\Psi_{n,m-1}
+e^{-2\pi i\phi n}\Psi_{n+1,m}) \nonumber \\ 
-t_c(e^{-2\pi i\phi (n-\frac{1}{2})}\Psi_{n-1,m+1}
+e^{2\pi i\phi (n+\frac{1}{2})}\Psi_{n+1,m}) =E\Psi_{n,m}. 
\end{eqnarray}
We take the form of the wavefunction $\Psi_{n,m}=e^{ik_y m}\Psi_n$, 
then the  Schr\"odinger equation becomes 
\begin{eqnarray}
-(t_a+t_ce^{-2\pi i \phi(n-\frac{1}{2})+ik_y})\Psi_{n-1} 
-(t_a+t_ce^{2\pi i\phi(n+\frac{1}{2}-ik_y)})\Psi_{n+1} 
-2t_b\cos(2\pi\phi n+k_y)\Psi_n =E\Psi_n.  
\label{eq;triharper}
\end{eqnarray}

When $\phi=\frac{p}{q}$ ($p$ and $q$ are coprime integers), (\ref{eq;triharper}) 
is periodic with period $q$. The Bloch theorem tells 
that one can put $\Psi_n = \exp(ik_x n)\psi_n$ where $\psi_n$ 
satisfies $\psi_n=\psi_{n+q}$, which implies that 
$\Psi_{n+q}=e^{ik_x q}\Psi_n$. Thus, if we introduce 
a row vector $\Psi=(\Psi_1,\Psi_2,\cdots,\Psi_{q-1},\Psi_q )^t$ ($t$ 
means the transpose of a matrix) and 
$a_n(k_y)=t_a+t_c\exp(2\pi i \frac{p}{q}(n+\frac{1}{2})-ik_y)$ and
 $b_n(k_y)=2t_b\cos(2\pi\frac{p}{q}n+k_y)$, 
(\ref{eq;triharper}) is reduced to 
an eigenvalue problem of a finite size matrix 
\begin{eqnarray}
 H_q(k_x,k_y) =   \left(
\begin{array}{@{\,}cccccccc@{\,}}
        b_{1}(k_y) &
        a_{1}(k_y) & 
        0 & \cdots  &  &  &  0 &
        e^{ik_x q} a_{0}(k_y)^{*} \\
        a_{1}(k_y)^{*} &
        b_{2}(k_y)&
        a_{2}(k_y) &
        0 & \cdots &  & & 0 \\
        0 &     
        a_{2}(k_y)^{*} &
        b_{3}(k_y) &
        a_{3}(k_y) &
        0 & \cdots &  & 0 \\
          & 0 &
        a_{3}(k_y)^{*}  &
        b_{4}(k_y) &
        a_{4}(k_y) &
        0 & \ldots & 0 \\
        \vdots & \vdots & \ddots &
        \ddots & \ddots & \ddots &
        \ddots & \vdots \\
        0 & & \cdots & 0 & 
        a_{q-3}(k_y)^{*} &
        b_{q-2}(k_y) &
        a_{q-2}(k_y) &
        0 \\
        0 & & &\cdots  & 0 & 
        a_{q-2}(k_y)^{*} &
        b_{q-1}(k_y) &
        a_{q-1}(k_y)
        \\
        e^{-ik_xq} a_{0}(k_y)  & 
        0 & &  &\cdots & 0 & 
        a_{q-1}(k_y)^{*}&
        b_{q}(k_y) \\
\end{array}
\right). 
\label{eq;matrix}
\end{eqnarray}
Also, in terms of $\psi_n$,  
(\ref{eq;triharper}) becomes 
\begin{eqnarray}
-e^{-ik_x}(t_a+t_ce^{-2\pi i\phi(n-\frac{1}{2})+ik_y})\psi_{n-1} 
-e^{ik_x}(t_a+t_ce^{2\pi i \phi(n+\frac{1}{2})-ik_y})\psi_{n+1} 
-2t_b\cos(2\pi\phi n+k_y)\psi_n =E\psi_n.  
\label{eq;triharpera}
\end{eqnarray}
At $t_c=0$, this is reduced to the Harper equation.  
We define   $\lambda \equiv 2\frac{t_b}{t_a}$ 
and $\mu \equiv \frac{t_c}{t_a}$   then (\ref{eq;triharper}) becomes  
\begin{eqnarray}
-\left[1+\mu e^{-2\pi i \frac{p}{q}(n-\frac{1}{2})+ik_y}\right]\Psi_{n-1} 
-\left[1+\mu e^{2\pi i \frac{p}{q}(n+\frac{1}{2}-ik_y)}\right]\Psi_{n+1} 
-\lambda \cos\left(2\pi\frac{p}{q} n+k_y\right)\Psi_n =E\Psi_n.  
\label{eq;triharperlambdamu}
\end{eqnarray}

\subsection{Hamiltonian in momentum space}
We denote $\boldsymbol{k}=(k_x,k_y)$. 
The electron annihilation operator $c(\boldsymbol{k})$ 
in momentum space is 
\begin{eqnarray}
c_{n,m} =\frac{1}{(2\pi)^2}
\int^{\pi}_{-\pi} dk_x \int^{\pi}_{-\pi} dk_y \exp[ik_x n+ik_y m]
c(\boldsymbol{k}). 
\end{eqnarray}
The commutation relation for $c(\boldsymbol{k}), c^{\dagger}(\boldsymbol{k})$ is 
\begin{eqnarray}
\{c(\boldsymbol{k}),c^{\dagger}(\boldsymbol{k}')\} = (2\pi)^2\delta_{Z}(k_x-k_x')\delta_{Z}(k_y-k_y'), 
\end{eqnarray}
where $\delta_Z(k)=\sum_{n \in \boldsymbol{Z}} \delta(k+2\pi n)$.  
In terms of $c(\boldsymbol{k}), c^{\dagger}(\boldsymbol{k})$, 
the tight-binding Hamiltonian (\ref{eq;model}) is 
\begin{eqnarray}
H= \frac{1}{(2\pi)^2} \int^{\pi}_{-\pi} dk_x 
\int^{\pi}_{-\pi} dk_y  H(\boldsymbol{k}),  
\end{eqnarray}
with 
\begin{eqnarray}
H(\boldsymbol{k})=&&
 -2t_a \cos k_x c^{\dagger}(\boldsymbol{k})c(\boldsymbol{k})  \nonumber \\ 
&&-(t_b e^{-ik_y}+t_c e^{-ik_x+ik_y-i\pi \phi} )
c^{\dagger}(k_x+2\pi\phi,k_y)c(k_x,k_y) \nonumber \\
&&- 
(t_be^{ik_y}+t_ce^{ik_x-ik_y-i\pi \phi})c^{\dagger}(k_x-2\pi\phi,k_y)c(k_x,k_y).
\label{eq;hamiltoniank}
\end{eqnarray}
When $\phi=p/q$, since $k_x$ couples only to $k_x \pm 2\pi \phi$, we write 
$k_x$ as $k_x^{0}+2\pi\phi j$  with $j \in \boldsymbol{Z}$. 
Here $k^{0}_x$ is in the 
magnetic Brillouin zone  
\begin{eqnarray}
-\frac{\pi}{q} \leq k^{0}_x \leq \frac{\pi}{q}. 
\end{eqnarray}
The Hamiltonian $H(\boldsymbol{k})$ acts on the Hilbert space 
 spanned by 
\begin{eqnarray}
\ket{\Psi} = \sum_{=j}^{q} \widetilde{\psi}_j c^{\dagger}(k_x^{0}+2\pi\phi j,k_y)\ket{0}, 
\end{eqnarray}
with $\widetilde{\psi}_{j+q}=\widetilde{\psi}_{j}$.  
The Schr\"odinger equation 
$H\ket{\Psi}=E\ket{\Psi}$ is 
\begin{eqnarray}
-(t_be^{-ik_y}+t_ce^{-ik^{0}_x+ik_y-2\pi i \phi(j -\frac{1}{2})})\widetilde{\psi}_{j-1}  
-(t_be^{ik_y}+t_ce^{ik^{0}_x-ik_y+2\pi i\phi(j+\frac{1}{2})})\widetilde{\psi}_{j+1}  
-2t_a \cos (2\pi\phi j+k^{0}_x) \widetilde{\psi}_j =E\widetilde{\psi}_j. 
\nonumber \\ 
\label{eq;triharperk}
\end{eqnarray}
In (\ref{eq;triharper}) and (\ref{eq;triharperk}),  
the terms $H_a$ and  $H_b$  are diagonal respectively. 
  We can also diagonalize $H_c$ which is proportional to $t_c$
by changing the gauge. For example, we take the gauge
\begin{eqnarray}
A_{n+1,m;n,m} =2\pi\phi (n+\frac{1}{2}),
\hspace{3mm}     A_{n,m+1;n,m} =2\pi \phi n, \hspace{3mm} {\rm and}
\hspace{3mm} A_{n,m+1;n+1,m} =0.  
\label{t_c-gauge} 
\end{eqnarray}
The gauge transformation which transforms from (\ref{t_a-gauge}) to 
(\ref{t_c-gauge}) is given by 
\begin{eqnarray}
c_{n,m} &\rightarrow& c_{n,m} \exp(if_{n}) \nonumber \\ 
c^{\dagger}_{n,m} &\rightarrow& c^{\dagger}_{n,m} \exp(-if_{n}) \nonumber \\ 
A_{n,m;n',m'} &\rightarrow& A_{n,m;n',m'} + f_{n,m} - f_{n',m'}, \nonumber \\ 
f_n &=&\phi n(n-1). 
\end{eqnarray}
In this gauge, the Hamiltonian in momentum space is 
\begin{eqnarray}
H= \frac{1}{(2\pi)^2} \int^{\pi}_{-\pi} dk_x 
\int^{\pi}_{-\pi} dk_y  H'(\boldsymbol{k}),  
\end{eqnarray}
with 
\begin{eqnarray}
H'(\boldsymbol{k})=&&
 -2t_c \cos (k_x-k_y) c^{\dagger}(\boldsymbol{k})c(\boldsymbol{k})  
\nonumber \\ 
&&-(t_b e^{-ik_y}+t_a e^{ik_x+\pi i \phi} )
c^{\dagger}(k_x+2\pi\phi,k_y)c(k_x,k_y) \nonumber \\
&&- 
(t_be^{ik_y}+t_ae^{-ik_x+\pi i \phi})c^{\dagger}(k_x-2\pi\phi,k_y)c(k_x,k_y) 
\label{eq;hamiltonianck}
\end{eqnarray}
From (\ref{eq;hamiltonianck}), we get the Schr\"odinger equation : 
\begin{eqnarray}
-(t_be^{ik_y}+t_ae^{-ik^{0}_x-2\pi i \phi (\ell -\frac{1}{2})})\widehat{\psi}_{\ell-1}  
-(t_be^{-ik_y}+t_ae^{ik^{0}_x+2\pi i \phi(\ell+\frac{1}{2})})\widehat{\psi}_{\ell+1}  
-2t_c \cos (2\pi\phi \ell +k^{0}_x-k_y) \widehat{\psi_\ell} =E\widehat{\psi}_\ell. 
\label{eq;triharperc}
\end{eqnarray}
Apparently, if one exchange $k_x^{0}$  by 
$k_x^{0}-k_y$ and $t_a$ by $t_c$
in (\ref{eq;triharperk}), we get (\ref{eq;triharperc}).  
This is due to the symmetry of the triangular lattice.

\subsection{Duality}
At $t_c = 0$, it is known that (\ref{eq;triharpera}) 
has the duality of Aubry and Andr\'{e} \cite{aubry}
who showed the existence of a transition 
between localized and extended states of $\psi_j$  when 
$\phi$ is an irrational number.  When $\lambda > 2$, 
the states are all localized, and when $\lambda < 2$, the states are 
all extended. At $\lambda=2$, all the states are critical.  

In the present case, take $\phi=\frac{p}{q}$  and write  
\begin{eqnarray}
\psi_n = \sum_{l=0}^{q-1} e^{2\pi \phi nl} f_l. 
\label{eq;fourier}
\end{eqnarray}
and substitute it into (\ref{eq;triharpera}), then
\begin{eqnarray}
-(t_be^{ik_y}+t_c e^{ik_x-ik_y+2\pi i\phi(l-\frac{1}{2})})f_{l-1} 
-(t_be^{-ik_y}+t_c e^{-ik_x+ik_y-2\pi i\phi(l+\frac{1}{2})})f_{l+1} 
-2t_a\cos(2\pi \phi l+k_x)f_l =Ef_l. 
\label{eq;dual}
\end{eqnarray}
When $t_c=0$, (\ref{eq;dual}) becomes (\ref{eq;triharperk}) by 
substituting $k_x \rightarrow k_x^{0}$ and $ k_y \rightarrow -k_y$ and 
$\lambda \rightarrow \frac{4}{\lambda}$. 
This is just the Aubry-Andr\'{e} duality when we 
take the incommensurate limit of $\phi$.  However, when $t_c \neq 0$, 
(\ref{eq;dual}) and (\ref{eq;triharperk}) are not 
transformed by  (\ref{eq;fourier}) due to the term proportional 
to $t_c$. 

Because of  the symmetry of the triangular lattice, we can 
consider duality involving $t_c$ by putting $t_a$ or $t_b$ to be zero. 
Let us consider small $t_a$ limit.  In  the limit, 
(\ref{eq;triharperc}) has the Aubry-Andr\'{e} duality for exchanging 
$t_b$ and $t_c$.  This implies that  there is a  duality between  
$\lambda$ and $\mu$ for small $t_a$ limit. 
It relates a state at $(\lambda,\mu)$ to the one at 
$(2\mu,\frac{\lambda}{2})$ by the transformation (\ref{eq;fourier}). 
Thus the phase diagram in $(\lambda,\mu)$ should 
have a localization transitions on the line $\lambda=\mu$ for 
small $t_a$ limit i.e. large $\lambda$ and  $\mu$.

\subsection{Characteristic Polynomial}
When $\phi=p/q$ is rational, (\ref{eq;triharper}) 
is reduced to the eigenvalue problem of the matrix (\ref{eq;matrix}).  
The eigenvalues are determined by the zeroes of 
the characteristic polynomial 
\begin{eqnarray}
P(E)=\det(E-H_q(k_x,k_y)). 
\end{eqnarray}
which has been studied  previously 
\cite{thouless,hat-koh,han-thouless}. In Ref.\cite{han-thouless}, 
it was shown that $P(E)$ can be written in terms of 
Chebyshev polynomial of order $q$. In the present case, 
the characteristic polynomial takes a simple form as follows: 
\begin{eqnarray}
P(E) &=& P_0(E)-Q(k_x,k_y)  \\ 
Q(k_x,k_y)&=&(-1)^{q}4
(t_a^{q}\cos qk_x +t_b^{q}\cos qk_y)+(-1)^{p}t_c^{q}\cos q(k_x-k_y),
\end{eqnarray}
where $P_0(E)$ is independent of $k_x$ and $k_y$.   
The energy bands are determined by the zeroes of the 
polynomial $P(E)$ as we vary $k_x$ and $k_y$. 
Especially, the edges of energy bands are determined by 
the minimum and the maximum of the function $Q(k_x,k_y)$. 
When $k_y=0$, they are given by $k_x=0, \pi$. 

In Ref.\cite{han-thouless}, the total band width $W$ of 
the triangular lattice is 
estimated when $t_b > t_a > t_c$ as 
\begin{equation}
W \sim (t_b-t_a) g\left(q\frac{t_b-t_a}{t_b} \right), 
\label{eq;totalband}
\end{equation}
where the scaling function $g(x)$ behaves as $\frac{9.3299}{x}$ 
when its argument is small. Since $t_c$  does not 
enter the argument, the scaling of the total band width 
of the triangular lattice is the same  as  the square lattice. 
This suggests that the universality class of 
the scaling property of the spectral measure of 
the triangular lattice is the same as that of the square 
lattice.

\section{\label{section;chara}Classical Orbits and Phase Diagram}
\subsection{Classical Orbits}

\begin{figure}
\includegraphics[scale=1.0]{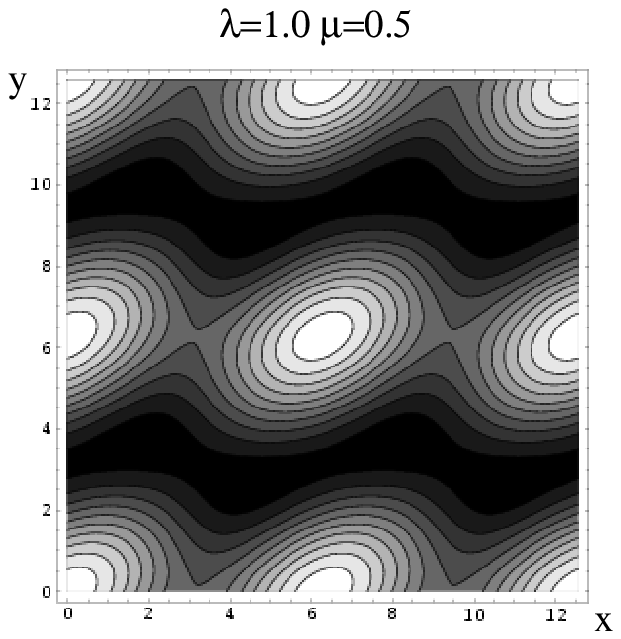}
\includegraphics[scale=1.0]{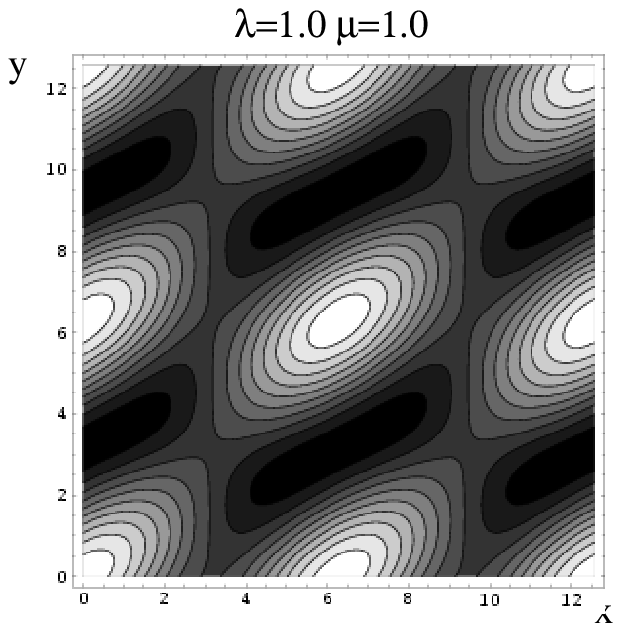}
\includegraphics[scale=1.0]{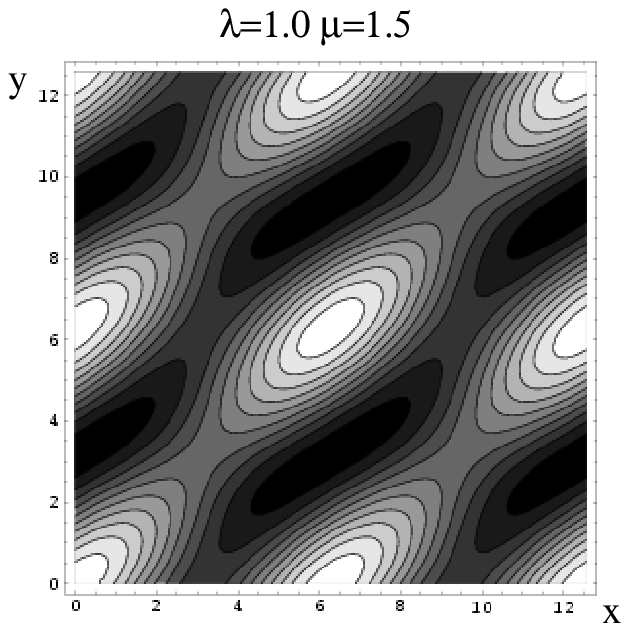}
   \caption{ Contour plots of the classical orbits 
    for 
$(\lambda , \mu) = (1.0 , 0.5),(1.0,1.0)$ and $(1.0,0.5)$. 
  }
   \label{fig;contour-1.0-1.0}
\end{figure}

\begin{figure}  
\includegraphics[scale=1.0]{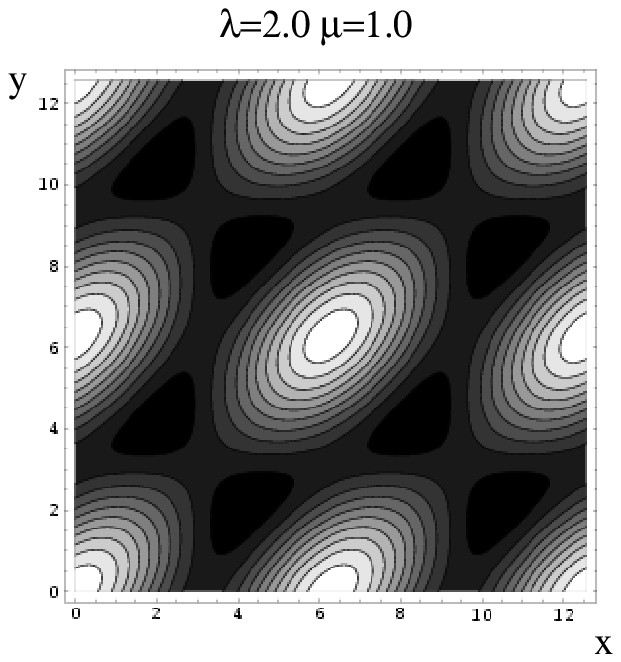}
\includegraphics[scale=1.0]{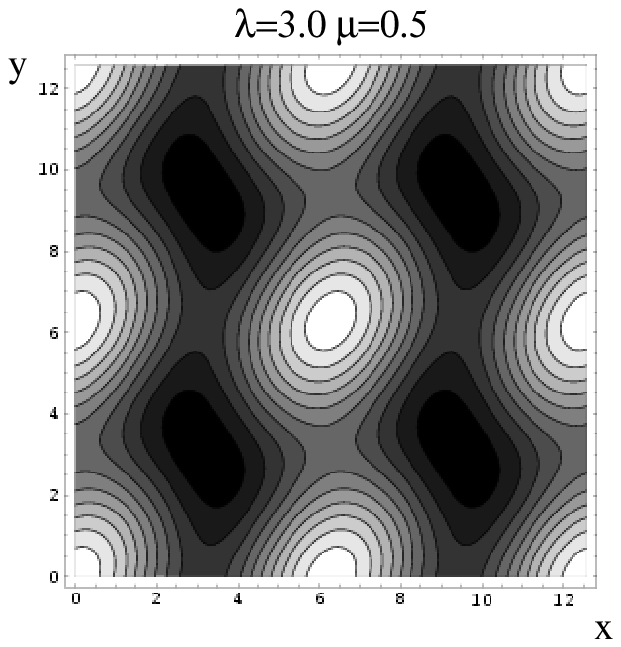}
 \caption{ Contour plots of the classical orbits 
    for 
$(\lambda , \mu) = (2.0,1.0)$ and $(3.0,0.5)$. 
   }
   \label{fig;contour-2.0-1.0}
\end{figure}

The Hamiltonian (\ref{eq;model}) consists of  three terms $H_a$, $H_b$ and $H_c$
which are noncommutative each other.
They are diagonalized in different bases 
as in (\ref{eq;triharper}), (\ref{eq;triharperk}) and 
(\ref{eq;triharperc}). 
The ``classical'' Hamiltonian is thus 
\begin{eqnarray}
H_{\rm classical} = 2t_a \cos k_x+ 2t_b \cos k_y +2t_c \cos (k_x-k_y). 
\label{eq;classical}
\end{eqnarray}
In a magnetic field, $k_y$ is canonically conjugate to $k_x$ and vice versa. 
Thus to analyze classical orbits, we replace  $k_y$ by $x$ and 
$k_x$ by $y$.  Setting  $t_a=1$, we plot the contour of 
\begin{eqnarray}
H_{\rm classical} = \cos y + \frac{\lambda}{2} \cos x 
+ \mu \cos (y-x)  
\label{eq;classical2}
\end{eqnarray}
in {\bf Fig. \ref{fig;contour-1.0-1.0}} for $(\lambda,\mu)= (1.0,0.5),(1.0,1.0)$ and $(1.0,1.5) $, and in {\bf Fig.\ref{fig;contour-2.0-1.0}}  for $(\lambda,\mu)= (2.0,1.0)$ and $(3.0,1.5)$.
In {\bf Fig.\ref{fig;contour-1.0-1.0}}, 
 we see that all the contours for  $(\lambda,\mu)=(1.0,0.5) $ and $ (1.0,1.5)$ are 
extended  in the $x$-direction  and  localized in the $y$-direction while, 
for $(\lambda,\mu )=(1.0,1.0)$, there is a separatrix which is 
extended in both directions.  
We also see  in {\bf Fig.\ref{fig;contour-2.0-1.0}} that 
the contours for $(\lambda,\mu)=(3.0,0.5)$ are 
extended  in $x$-direction  and  localized in $y$-direction, while
there is a separatrix for $(\lambda,\mu)=(2.0,1.0)$.

\subsection{Phase Diagram for irrational $\phi$}
\begin{figure}
  \begin{center}
    \epsfxsize=8cm
    \epsfbox{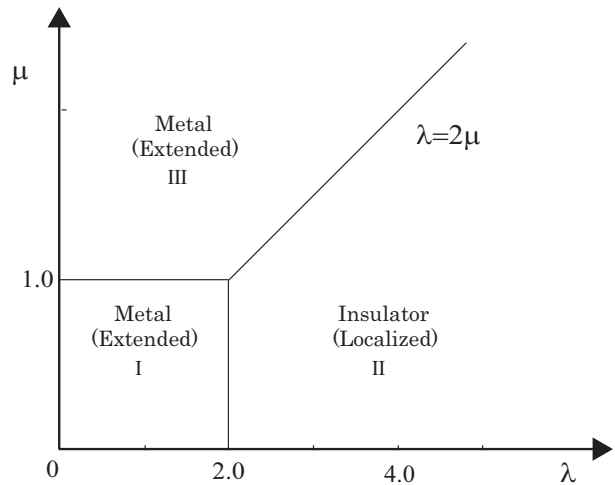}
    \caption{  
     Phase diagram.
      In region I and III
     the wavefunctions (spectra) are extended 
      (absolutely continuous),
      and in region II
      the wavefunctions are localized (pure points).
      On the three boundary lines, 
  the wavefunctions (spectra) are critical (singular continuous).
    }
    \label{fig;phase_diagram}
  \end{center}
\end{figure}

From behaviors of  the classical orbits shown in  the previous section, 
we may deduce the phase diagram of the equation 
\begin{eqnarray}
-\left[1+\mu e^{-2\pi i \phi(n-\frac{1}{2})+ik_y}\right]\Psi_{n-1} 
-\left[1+\mu e^{2\pi i \phi(n+\frac{1}{2}-ik_y)}\right]\Psi_{n+1} 
-\lambda \cos\left(2\pi\phi n+k_y\right)\Psi_n =E\Psi_n.  
\label{eq;triharperlambdamu2}
\end{eqnarray}
for irrational limit of $\phi=\frac{p}{q}$. 
The phase diagram is shown in 
{\bf Fig.\hspace{-.2cm} \ref{fig;phase_diagram}}.
One intriguing aspect is the 
transitions between phase I and phase III
which are the transitions between metals.  
Indeed a transition in the quantum case is not 
characterized by an appearance of a separatrix at a certain 
energy. For the Harper model, it is known that 
metal-insulator transitions occur for whole energies 
at $\lambda=2$. This is generalized to the triangular lattice model 
we consider.

As an example of an incommensurate limit, 
in the sections hereafter, we perform numerical scaling analysis 
for the energy spectra and the critical wavefunctions 
when $\phi=\frac{p}{q}$ approaches the inverse of the golden mean 
$\frac{1}{\tau}=\frac{\sqrt{5}-1}{2}$.  A standard sequence 
which corresponds to the continued fraction expansion 
of $\frac{1}{\tau}$ is the Fibonacci series $F_n$, 
which is defined by $F_0=F_1=1$, $F_n=F_{n-1}+F_{n-2}$. 
$F_n$ behaves $\sim \tau^{n}$ for large $n$. 
By taking $p=F_{n-1},q=F_n$, $\phi=\frac{p}{q}$ approaches 
$\frac{1}{\tau}$.  $F_n$ is called a Fibonacci number 
and $n$ is referred to Fibonacci index.

To take this incommensurate limit of (\ref{eq;triharperlambdamu2}), 
the  off-diagonal terms in   (\ref{eq;triharperlambdamu2}) need 
 a special care. Namely, when $\mu=1$, these terms can be zero 
 if the exponential becomes $-1$.  For the sequence above, 
 this actually happens when $q=F_n$ with $n=3\ell+1$  for some integer $\ell$. 
 In that case, the energy spectrum has  no dependence on $k_x$, 
 and the dispersion relation is flat.

\section{\label{section;level}Level Statistics}

Consider (\ref{eq;triharperlambdamu2})  when $k_y=0$. 
On the critical lines, the spectral measure
 and the wavefunctions are expected to show characteristic 
  behaviors of criticality. 
See {\bf Fig.\ref{fig;phase_diagram}}.
In order to obtain the distributions of the band widths,  the $q \times q$ matrices  (\ref{eq;matrix}) are diagonalized. The normalizations are 
\begin{eqnarray}
\int^{\infty}_0 P_B(w) dw &=&1 
\nonumber\\
\langle w \rangle = \int^{\infty}_0 w P_B(w)dw &=&1.
\end{eqnarray}
Similarly the distributions of the gaps $P_G(s)$
are obtained and normalized by

\begin{eqnarray}
\int^{\infty}_0 P_G(s) ds &=&1 
\nonumber\\
\langle s \rangle = \int^{\infty}_0 s P_G(s)ds &=&1.
\end{eqnarray}
As we discussed in Sec.\ref{section;chara}, the edges of the energy bands 
are found at $k_x=0,\pi$ and $k_y=0,\pi$.  
Thus, to study  the measure of the spectrum of (\ref{eq;triharperlambdamu2}), 
it is sufficient to study those points in the Brillouin zone.    
When $\phi=\frac{p}{q}$ is a rational number, the problem is reduced to the 
eigenvalue problem of the finite size matrix (\ref{eq;matrix}). 
Furthermore, when $q$ is odd and  $k_x=0,\pi$ and $k_y=0,\pi$, 
the matrix (\ref{eq;matrix}) reduces to a tridiagonal form by 
the symmetric and antisymmetric eigenstates \cite{thouless}.

For $\mu=1$ with $q=F_n$ and $n=3\ell+1$,  as we noted above, 
 the  hopping term becomes zero at a  bond 
and all the band has  zero width. Thus we study only the case of
 $q=F_n$ with $n=3\ell$ when $\mu=1$.

\subsection{ Distributions of Band Widths }

\begin{figure}
\includegraphics[scale=0.6,angle=0]{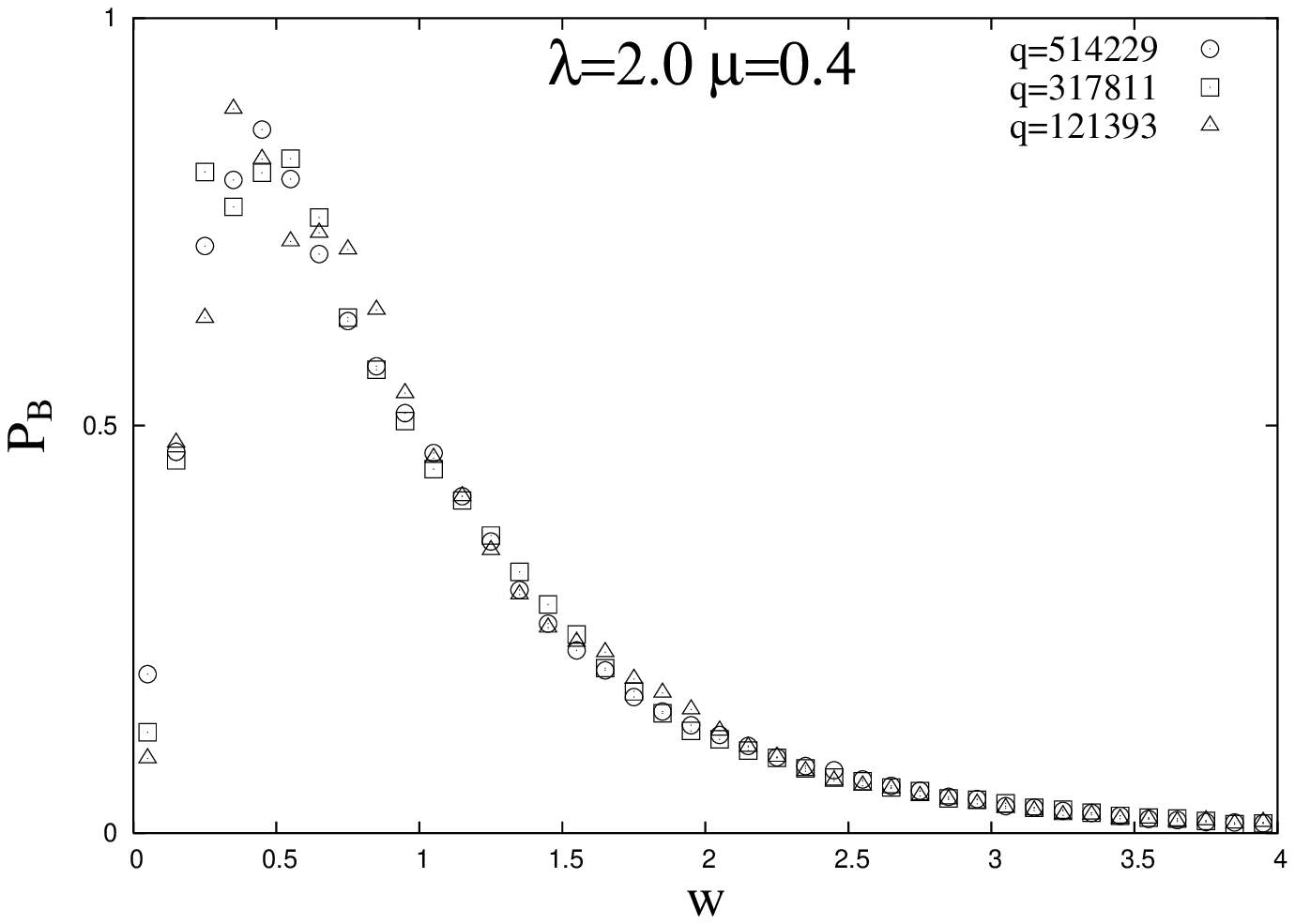}
   \caption{ Distributions of the band widths for 
  $(\lambda , \mu) = (2.0,0.4)$.
   }
   \label{fig;bw-fig-2.0-0.4}
\end{figure}

\begin{figure}
\includegraphics[scale=0.6,angle=0]{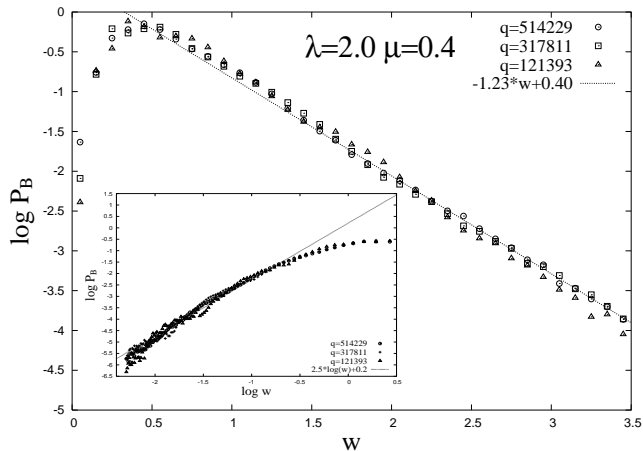}
   \caption{ Semi-log plots of  the distributions of the band widths
    for  $(\lambda , \mu) = (2.0,0.4)$.  Inset: 
Log-log plots of  the distributions of the band widths
    for  $(\lambda , \mu) = (2.0,0.4)$.   As seen in the inset, 
the convergence of the distributions near zero at large $n$ 
is relatively slow. 
For each value of $(\lambda,\mu)$, 
we choose a stable part of the distributions to obtain the 
exponents. This procedure potentially underestimates the values. 
   }
   \label{fig;bw-figlog-2.0-0.4}
\end{figure}


Consider the distributions of the band widths
along the line $\lambda = 2$ with $\mu=0.2,0.4,0.6$ and $0.8$. 
In {\bf Fig. \ref{fig;bw-fig-2.0-0.4}}, $P_B(w)$ at $( \lambda , \mu )=( 2.0 , 0.4 )$
for $q=F_n$ with  $n=25, 27$ and $ 28$ are plotted.
It shows  convergence to a limit,
indicating the existence of a limit of the 
distributions of the gaps for the incommensurate flux $\varphi$.
For $0 \leq \mu < 1$
the distributions depend on $\mu$. 
The semi-log plots 
of $P_B(w)$ is shown in {\bf Fig.   \ref{fig;bw-figlog-2.0-0.4}}. One sees the linear behaviors 
for large $w$, implying an asymptotic form 
\begin{equation}
  P_B(w) \sim e^{-\gamma w},\quad {\rm as}\quad  w \rightarrow \infty, 
  \label{tail}
\end{equation} 
where $\gamma >0.$ The optimized values of  $\gamma$  are shown
in {\bf Table \ref{table;level}} for several $\mu$'s. 
The inset of {\bf Fig.\ref{fig;bw-figlog-2.0-0.4}}
shows $P_B(w)$ near the origin which 
indicates that the distributions of the band widths $P_B(w)$  are  zero at the origin with  a power law decay.
To characterize this behavior, we make an ansatz 
\begin{equation}
  P_B(w) \sim w^{\beta}, \quad {\rm as } \quad w \rightarrow 0.
  \label{origin}
\end{equation}
where $\beta >0$.   
The optimized values of  $\beta$ are shown 
in {\bf Table \ref{table;level}}.    
One sees that $\beta$  becomes smaller as approaching to 
 $\mu=1$.

\begin{figure}
\includegraphics[scale=0.6,angle=0]{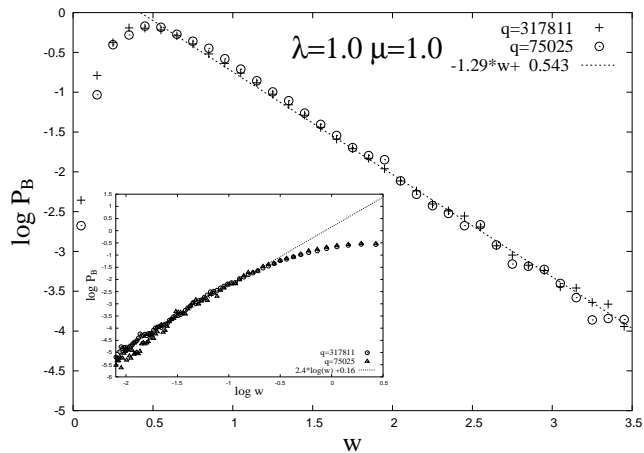}
   \caption{ Semi-log plots of  the distributions of the band widths
   for $(\lambda , \mu) = (1.0,1.0)$. Inset: 
  Log-log plots of  the distributions of the band widths
    for  $(\lambda , \mu) = (1.0,1.0)$.   }
   \label{fig;bw-figlog-1.0-1.0}
\end{figure}

Next, we investigate $P_B(w)$ on the other  lines $\mu=1$ and $\lambda=2\mu$.  
In {\bf Fig.   {\bf \ref{fig;bw-figlog-1.0-1.0} }}, 
the semi-log and the log-log 
plots of $P_B(w)$ are shown for $(\lambda,\mu)=(1.0,1.0)$. 
We find similar  type of  behaviors 
(\ref{tail}) and (\ref{origin})
for the $\lambda=2$ line. 
We also investigate the critical line $\lambda=2\mu$ and 
find similar type of behaviors.  We collect the values of 
$\beta$,$\gamma$ in  {\bf Table \ref{table;level}}.

These behaviors  of $P_B(w)$ on these lines 
are  consistent with  the behavior of $P_B(w)$ in other quasiperiodic 
model \cite{evangelou,takada,naka} thus gives a support for 
the phase diagram of {\bf Fig.\ref{fig;phase_diagram}}. 

\subsection{Distributions of Gaps}
\begin{figure}
\includegraphics[scale=0.3,angle=-90]{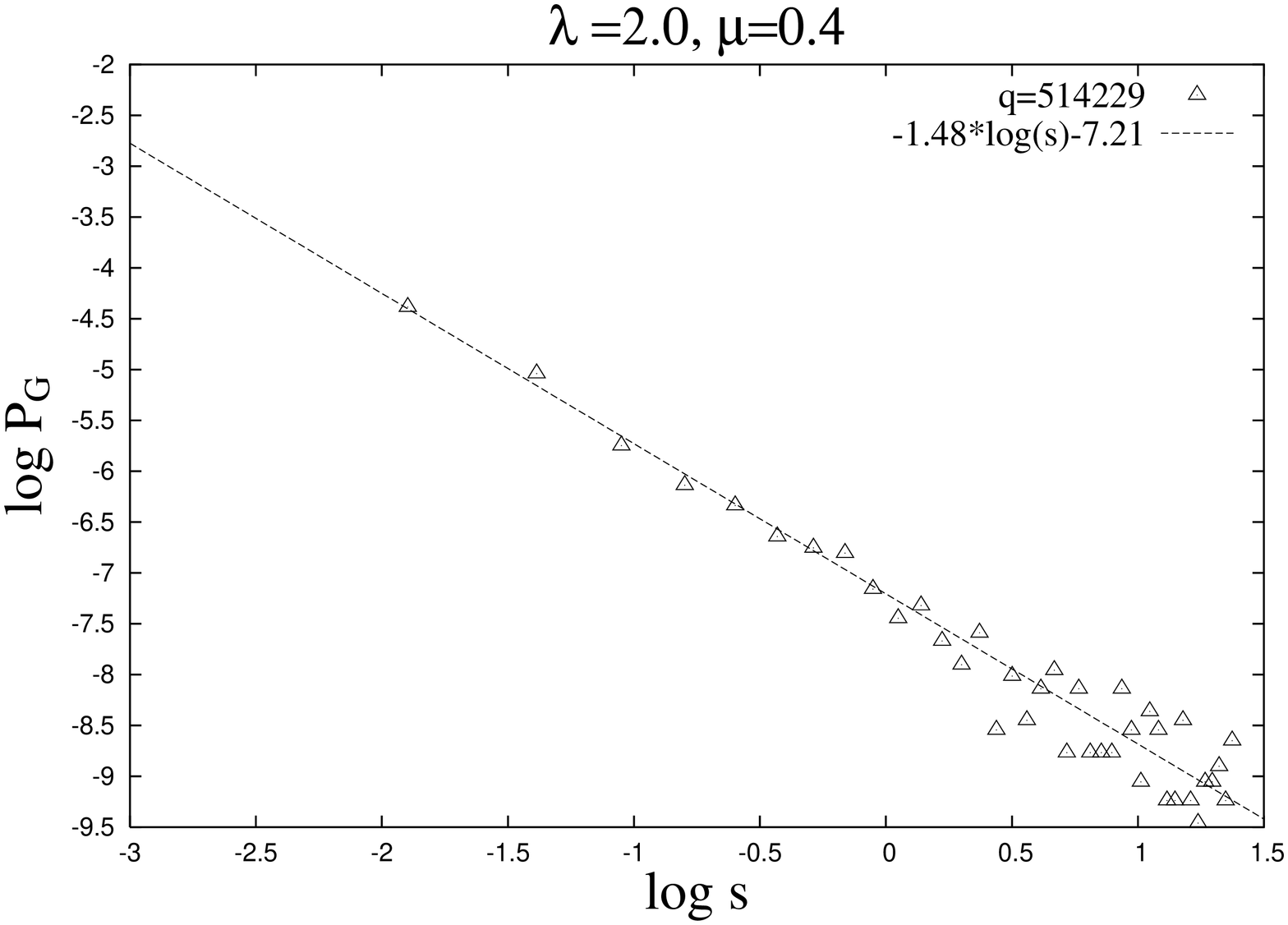}
\includegraphics[scale=0.3,angle=-90]{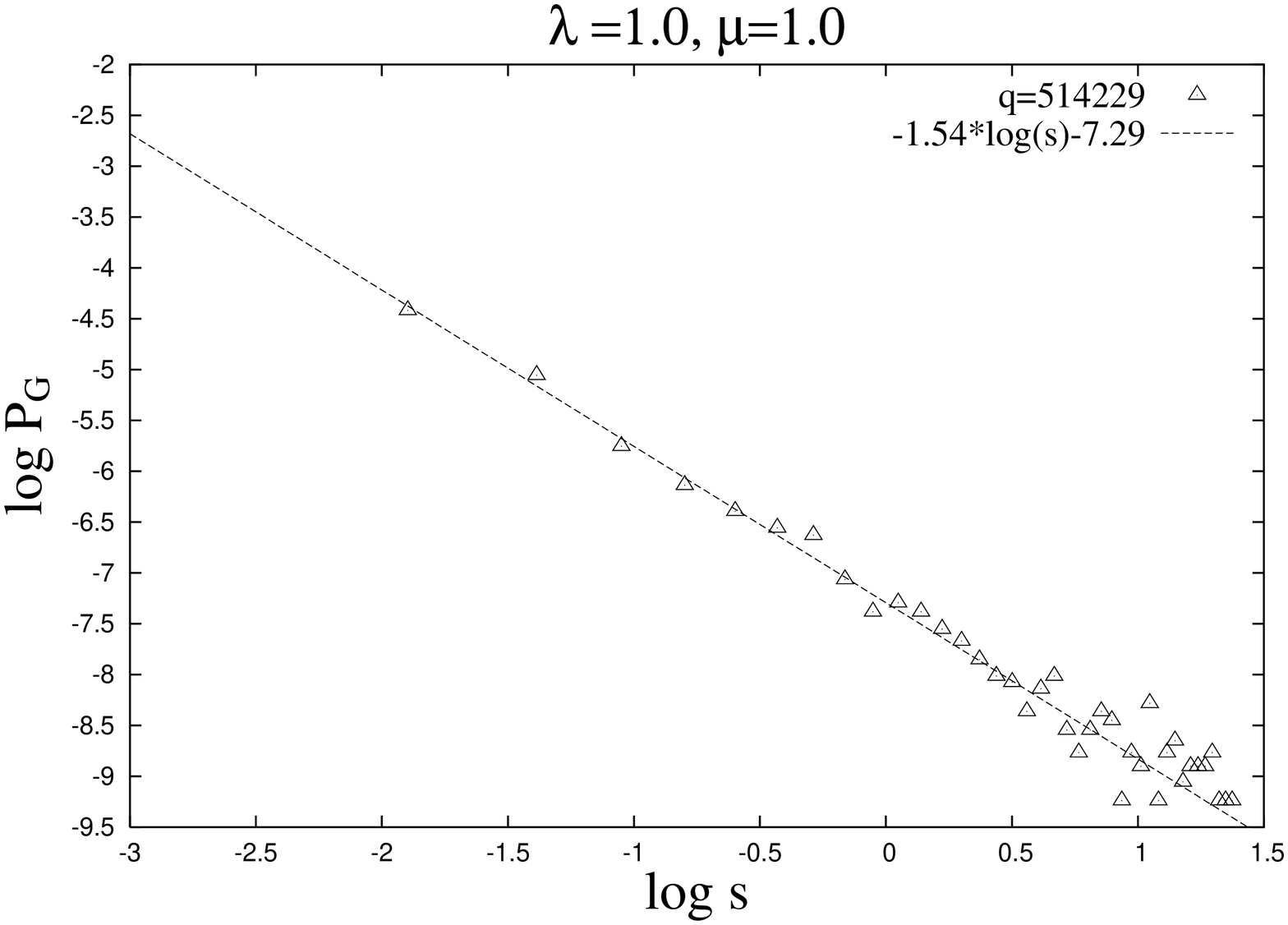}
\includegraphics[scale=0.3,angle=-90]{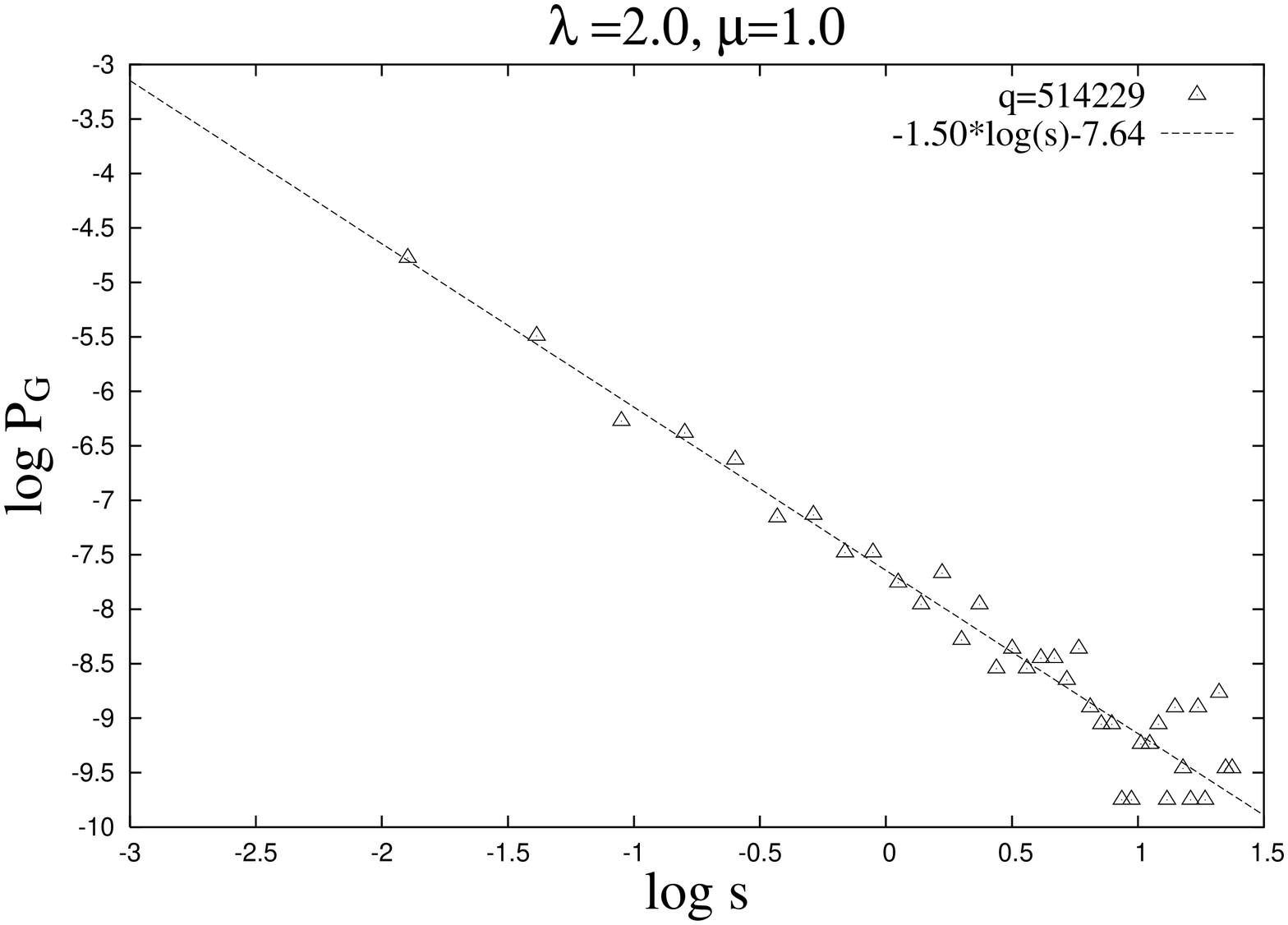}
   \caption{ Log-log plots of the distributions  of the gaps
    for  $(\lambda , \mu) = (2.0,0.4),(1.0 , 1.0)$. 
   }
   \label{fig;gap}
\end{figure}

The distribution of the gaps at $( \lambda , \mu) = ( 2.0 , 0.0 )$
has been known to follow  an inverse power law \cite{machigei}
which diverges at the origin 
\begin{equation}
P_G(s) \sim s^{-\delta}
\label{eq;gap_fit}
\end{equation}
with $\delta \sim 1.5$. 
In {\bf Fig. \ref{fig;gap}}, the distributions of the gaps
for $(\lambda,\mu)=(2.0,0.4), (1.0,1.0)$ and $(2.0,1.0) $ are 
shown. It is clear that 
$P_{G}(s)$ shows a power law of the inverse. 
The estimated value of $\delta$ is $\sim 1.5$ for these cases. 
We also investigate other points on the lines $\lambda=2, \mu=1$ 
and $\lambda=2\mu$ 
and find a similar behavior with $\delta \sim 1.5$ within 
statistical error. This behavior  of $P_G(s)$  shows 
that the spectra are singular continuous on these lines, 
gives a further support for 
the phase diagram {\bf Fig.\ref{fig;phase_diagram}}. 
Also  the value $\delta \sim 1.5$ seems to be  
 a characteristic quantity for this model.

\subsection{Bicritical Point}
\begin{figure}
\includegraphics[scale=0.3,angle=-90]{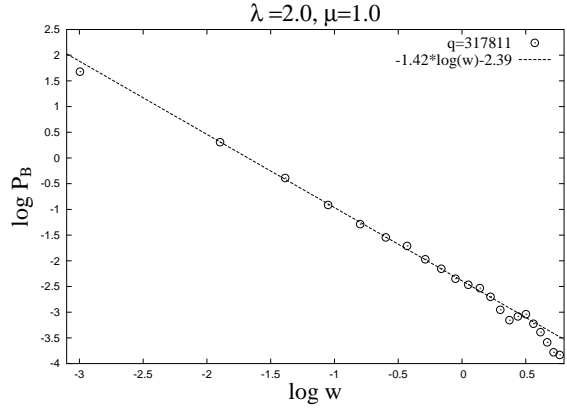}
   \caption{ Log-log plots of the distributions of the band widths
    for  $(\lambda , \mu) = (2.0,1.0)$. 
   }
   \label{fig;bw-2.0-1.0}
\end{figure}

\begin{figure}
\includegraphics[scale=0.3,angle=-90]{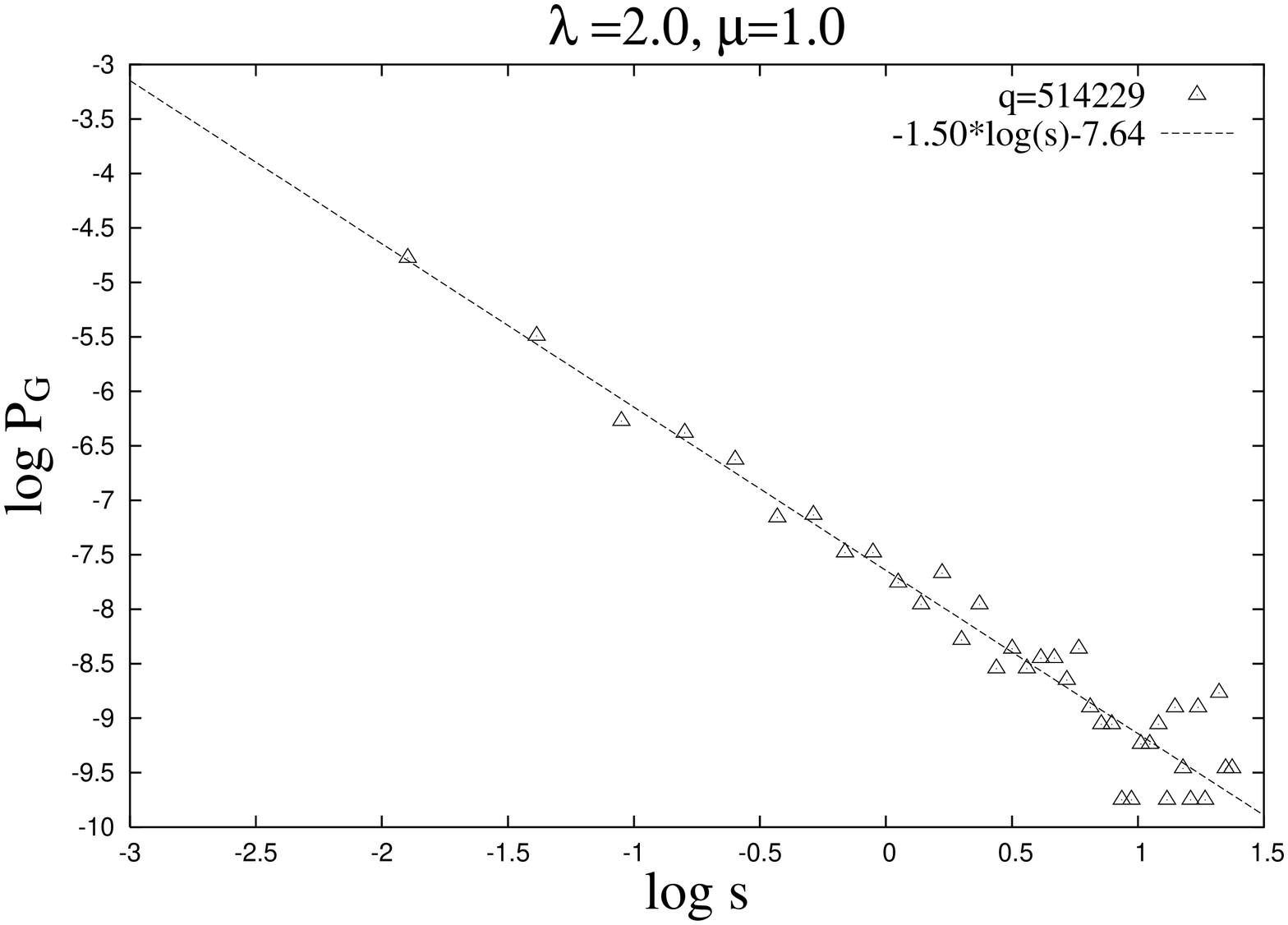}
   \caption{ Log-log plots of the distributions  of the gaps
    for  $(\lambda , \mu) = (2.0,1.0)$. 
   }
   \label{fig;gap-2.0-1.0}
\end{figure}

We also investigate the point  
$( \lambda , \mu) = ( 2.0 , 1.0 )$.  
The  result is shown in 
{\bf Fig.\ref{fig;bw-2.0-1.0}}. 
In sharp contrast to other points on the critical lines, 
it shows the inverse power law 
\begin{equation}
P_B(w) \sim w^{-\beta'}, 
\end{equation}
($\beta' >0$) for whole the range.  
We estimate  the exponent of the law as $\beta' \sim 1.4$. 
This implies that the spectrum at this point 
 is  a qualitively different fractal-like set. 
On the other hand,  the gap distribution is shown 
in {\bf Fig.\ref{fig;gap-2.0-1.0}}.  It is an 
inverse power law 
\begin{equation}
P_G(s) \sim s^{-\delta}, 
\end{equation}
($\delta>0$)
with the exponent $\delta \sim 1.5$,  
analogous to the ones found for 
 other critical points.  Thus the band width distribution 
 gives a finer characterization of the spectra than 
 the gap distribution.

\setlength{\arrayrulewidth}{0.8pt}
\begin{table}[tb]
\begin{tabular}{ccccc}     \hline
$\lambda$ & $\mu$ & $\beta $ 
& $\gamma$ & $\delta$  \\  \hline
2.0 & 0.0 & 2.5 & 1.4 & 1.5    \\
2.0 & 0.2 & 2.5 & 1.3   & 1.5     \\
2.0 & 0.4 & 2.5 & 1.2  & 1.5     \\
2.0 & 0.6 & 2.3 & 1.2  & 1.5     \\
2.0 & 0.8 & 2.1 & 0.9   & 1.5     \\
\hline
2.5 & 1.25& 2.1 & 0.9  &  1.5    \\
3.0 & 1.5 & 2.3 & 1.1  &   1.5   \\
4.0 & 2.0 & 2.4 & 1.2  &  1.5    \\
\hline
0.0 & 1.0 & 2.6 & 1.6  & 1.5    \\
0.5 & 1.0 & 2.6 & 1.5  & 1.5    \\
1.0 & 1.0 & 2.4 & 1.3  &  1.5    \\
1.5 & 1.0 & 2.2 & 1.0  &  1.5    \\

\end{tabular}
\caption{
Estimated exponents on the critical lines. 
For definitions of $\beta,\gamma$ and $\delta$,
see 
Eqs.(\ref{tail}), 
(\ref{origin}), and
(\ref{eq;gap_fit}) respectively.}
\label{table;level}
\end{table}

\section{\label{section;multi}Multifractal Analysis}

We apply the method of multifractal analysis \cite{halsey} 
to the spectra and the critical wavefunctions. In Ref.\cite{kohmoto}, 
the entropy function was introduced which reformulates 
the theory    along 
the way that standard statistical mechanics is formulated. 
We use it in our analysis. 

\subsection{Review of Multifractal Analysis}
We consider quantities $l_i$ and their probability measure $p_i$ of 
a fractal-like set.  
Though we shall only consider the cases where 
$l_i$ or $p_i$ is a constant, a general formulation is reviewed for 
convenience. 
It is natural to consider distributions of logarithm of $l_i$ 
\begin{equation}
\varepsilon_i = -\frac{\ln l_i}{n}, \quad i.e.\quad l_i=\exp(-n\varepsilon).
\label{kohmoto:eq2.1}
\end{equation}
As $n$ becomes large, $l_i$ approaches zero , but $\epsilon_i$ takes a 
finite nonzero value for critical points.
We introduce a scale index $\alpha_i$ as the exponent 
of $p_i$ measured by $l_i$ as 
\begin{equation}
p_i = l_i^{\alpha_i}, \hspace{5mm}
\alpha_i = -\frac{1}{\epsilon_i}\frac{1}{n} \ln p_i. 
\label{kohmoto:eq3.1a}
\end{equation}
We write the number of $l_i$ whose scale index lies between 
$\varepsilon$ and $\varepsilon+d\varepsilon$, $\alpha$ and 
$\alpha+d\alpha$ as $\Omega(\varepsilon,\alpha)d\varepsilon d\alpha$. 
We take an ansatz that $\Omega(\varepsilon,\alpha)$ has 
the following scaling form for large $n$
\begin{equation}
\Omega(\varepsilon,\alpha) = \exp[nQ(\varepsilon,\alpha)], 
\label{kohmoto:eq3.2}
\end{equation} 
where $Q(\varepsilon,\alpha)$ can be seen as a kind of 
 entropy function.

Following \cite{halsey,kohmoto}, we consider the generalized 
partition function 
\begin{eqnarray}
\Gamma(q,\beta) &=& \sum_i p_i^{q}l_i^{\beta}  \\ 
 &=& \sum_i \exp[-n\varepsilon_i(\alpha_iq+\beta)].  
\label{kohmoto:eq3.3}
\end{eqnarray}
The generalized free energy is 
\begin{eqnarray}
G(q,\beta) = \frac{1}{n} \ln \Gamma(q,\beta), 
\label{kohmoto:eq3.4}
\end{eqnarray}
Using $Q(\varepsilon,\alpha)$, (\ref{kohmoto:eq3.3}) is written  
\begin{eqnarray}
\Gamma(q,\beta) = \int d\varepsilon \int d\alpha 
\exp[n[Q(\varepsilon,\alpha)-(\alpha q+\beta)\varepsilon]].
\label{kohmoto:eq3.6}
\end{eqnarray}
For large $n$, the maximum of the exponent dominates 
the integral and gives 
\begin{eqnarray}
G(q,\beta) = Q(\vev{\varepsilon}, \vev{\alpha}) - (\vev{\alpha}q+\beta)
\vev{\varepsilon},
\label{kohmoto:eq3.7}
\end{eqnarray}
  where $\vev{\varepsilon}$ and $\vev{\alpha}$ give the maximum 
of $Q(\varepsilon,\alpha)-(\alpha q +\beta)\varepsilon$, so we have 
\begin{eqnarray}
\frac{\partial Q(\varepsilon,\alpha)}{\partial \varepsilon} 
|_{\varepsilon=\vev{\varepsilon},\alpha=\vev{\alpha}} = \vev{\alpha}+\beta.
\label{kohmoto:eq3.8}
\end{eqnarray}
and 
\begin{eqnarray}
\frac{\partial Q(\varepsilon,\alpha)}{\partial \alpha} = \vev{\varepsilon}q.
\label{kohmoto:eq3.9}
\end{eqnarray}
Thus $G(q,\beta)$ is obtained from $Q(\varepsilon, \alpha)$ using 
(\ref{kohmoto:eq3.7})(\ref{kohmoto:eq3.8}) and (\ref{kohmoto:eq3.9}). 
From (\ref{kohmoto:eq3.9}), 
the maximum of $Q(\varepsilon,\alpha)$ with respect to 
$\alpha$ occurs when $q=0$. 
On the other hand, once $G(q,\beta)$ is calculated, 
$\vev{\varepsilon},\vev{\alpha}$ and $Q(\vev{\epsilon},\vev{\alpha})$ 
are given by 
\begin{eqnarray}
\vev{\varepsilon} = -\frac{\partial}{\partial \beta}G(q,\beta),
\label{kohmoto:eq3.10} \hspace{5mm}
\vev{\alpha}\vev{\varepsilon} = - \frac{\partial}{\partial q} G(q,\beta),
\label{kohmoto:eq3.11}
\end{eqnarray}
and 
\begin{eqnarray}
Q(\vev{\varepsilon},\vev{\alpha}) = G(q,\beta) 
- q\frac{\partial G(q,\beta)}{\partial q}
- \beta\frac{\partial G(q,\beta)}{\partial \beta}. 
\label{kohmoto:eq3.12}
\end{eqnarray}
Since $\vev{\varepsilon}$ and $\vev{\alpha}$ are functions of $q$ and 
$\beta$, different regions with scaling indices $\varepsilon$ and 
$\alpha$ are explored by changing the values of the parameters $q$ and 
$\beta$. Thus $Q(\vev{\varepsilon},\vev{\alpha})$ is implicitly 
a function of $q$ and $\beta$. 


The limit of $G(q,\beta)$ for large $n$, may be obtained by
\begin{eqnarray}
G(q,\beta_c(q)) = 0,  
\label{kohmoto:eq3.19}
\end{eqnarray}
and $\beta_c(q)$ can be regarded as a set of generalized dimensions. The scaling index $\vev{\varepsilon}_c$ which corresponds to  $\beta_c(q)$ could be 
considered as being a representative for a particular value of $q$. 

From (\ref{kohmoto:eq3.7}) 
(\ref{kohmoto:eq3.8})  and (\ref{kohmoto:eq3.19}), we see that 
$Q(\vev{\varepsilon},\vev{\alpha})$ at the critical point satisfies the 
relation 
\begin{eqnarray}
Q(\vev{\varepsilon}_c,\vev{\alpha}_c) &=& 
\frac{\partial Q(\epsilon,\vev{\alpha}_c)}{\partial \varepsilon}|_{\varepsilon=\vev{\varepsilon}_c}\vev{\varepsilon}_c =  f(\vev{\alpha}_c)\vev{\varepsilon}_c,
\label{kohmoto:eq3.21}
\end{eqnarray}
where $f(\vev{\alpha}_c)$ is given by 
\begin{equation}
f(\vev{\alpha}_c) = 
\frac{\partial Q(\epsilon,\vev{\alpha}_c)}{\partial \varepsilon}|_{\varepsilon=\vev{\varepsilon}_c}.
\label{kohmoto:eq3.22}
\end{equation} 
By substituting (\ref{kohmoto:eq3.22}) into (\ref{kohmoto:eq3.8}) and
(\ref{kohmoto:eq3.9}), we obtain 
\begin{equation}
f(\vev{\alpha}_c) = \vev{\alpha}_c q+\beta_c(q)
\label{kohmoto:eq3.23}
\end{equation}
and 
\begin{equation}
\frac{df(\vev{\alpha}_c)}{d\alpha}|_{\alpha=\vev{\alpha}_c} 
= q,
\label{kohmoto:eq3.24}
\end{equation}
respectively.  And (\ref{kohmoto:eq3.23}) and (\ref{kohmoto:eq3.24}) give 
\begin{equation}
\vev{\alpha}_c = -\frac{d \beta_c(q)}{dq}.
\label{kohmoto:eq3.25}
\end{equation}
Thus once $\beta_c(q)$ is known by solving (\ref{kohmoto:eq3.19}), 
$\vev{\alpha}_c$  and
$f(\vev{\alpha}_c)$ are obtained from (\ref{kohmoto:eq3.23}) and (\ref{kohmoto:eq3.25}). 
In terms of $f(\alpha)$, the density function of 
$\varepsilon$ and $\alpha$ $\Omega(\varepsilon,\alpha)$ 
is written, using (\ref{kohmoto:eq3.2}) and (\ref{kohmoto:eq3.21}) as 
\begin{eqnarray}
\Omega(\vev{\varepsilon},\vev{\alpha}_c) =\exp[n\vev{\varepsilon}_cf(\vev{\alpha}_c)] = \vev{l}_c^{-f(\vev{\alpha}_c)}, 
\label{kohmoto:eq3.26}
\end{eqnarray}
where $\vev{l}_c = \exp(-n\vev{\varepsilon}_c)$ is a representative 
length, and $\vev{\varepsilon}_c$ and $\vev{\alpha}_c$ are 
functions of $q$ [see (\ref{kohmoto:eq3.10}) and (\ref{kohmoto:eq3.25})]. 
$f(\alpha)$ can be considered to be a set of generalized dimensions. 
In numerical approach, we calculate $f(\alpha)$ for a given fractal-like 
object for a finite $n$ and extrapolate it to the limit 
$n \rightarrow \infty$.   From  $G(q,\beta)$ (\ref{kohmoto:eq3.4}), $G_n(q,\beta)$ at 
finite $n$ should behave as $G_n(q,\beta) \sim G(q,\beta) + O(\frac{1}{n})$. 
Thus we should extrapolate $G_n(q,\beta)$ as  a function of $\frac{1}{n}$ 
and estimate the limit for $n \rightarrow \infty$.

We denote the support of $f(\alpha)$ by $[\alpha_{\rm min},\alpha_{\rm max}]$ 
 and the value of $\alpha$ which gives the maximum of $f(\alpha)$ 
by $\alpha_0$. 

\subsection{Spectrum}
In this section, 
scaling properties of energy spectra are analyzed by the multifractal analysis. 

\subsubsection{Multifractal Analysis of Spectrum} 
We apply the general formulation to   characterization of the
energy spectra. Take  the band widths as variables $l_i$, and let $p_i$ be
constants  
\begin{equation}
p_i=\frac{1}{F_n} \sim  \frac{1}{\tau^n}, 
\label{kohmoto:eq4.7} 
\end{equation}
where $\tau$ is the golden mean. As seen from  (\ref{kohmoto:eq3.4}), 
\begin{eqnarray}
G(q,\beta) = -q\ln \tau +F(\beta), \hspace{5mm} F(\beta) 
= G(0,\beta) = \frac{1}{n}\ln \Gamma(0,\beta)
\label{kohmoto:eq4.8}. 
\end{eqnarray}
For each $q$, the critical value $\beta$ is determined and 
vice versa.  
From (\ref{kohmoto:eq3.10}) and (\ref{kohmoto:eq4.8}), the 
scaling index $\vev{\varepsilon}$ is witten 
\begin{equation}
\vev{\varepsilon} = -\frac{\partial}{\partial \beta} G(q,\beta)
=-\frac{\partial}{\partial \beta} F(\beta). 
\label{kohmoto:eq4.9}
\end{equation}
Thus $\vev{\varepsilon}$ depends only on  $\beta$. 
From (\ref{kohmoto:eq3.11}) and (\ref{kohmoto:eq4.8}),
$\vev{\alpha}$ is related to $\vev{\varepsilon}$ by 
\begin{equation}
\vev{\alpha} = \ln \tau / \vev{\varepsilon},
\label{kohmoto:eq4.10}
\end{equation}
and $Q(\varepsilon,\alpha)$ is nonzero only for  $\alpha$ satisfying 
(\ref{kohmoto:eq4.10}), thus depends only on $\varepsilon$. 
We may put $Q(\varepsilon,\alpha)$ as $S(\varepsilon)$, and  
from (\ref{kohmoto:eq3.12}) and (\ref{kohmoto:eq3.21}) we get    
\begin{eqnarray}
S(\varepsilon)= F(\beta) +\varepsilon\beta=\varepsilon f(\alpha).
\label{kohmoto:eq4.11}
\end{eqnarray}
Then $f(\alpha)$ is calculated from the formula 
\begin{equation}
f(\alpha) = \frac{S(\varepsilon)}{\varepsilon}.
\end{equation}

\subsubsection{Numerical Results}
\begin{figure}
\includegraphics[scale=0.6,angle=-90]{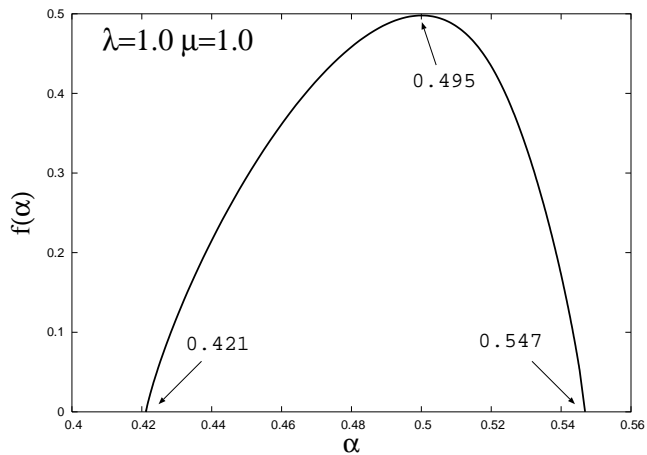}
   \caption{ $\alpha$-$f(\alpha)$ curves of the spectrum 
    for  $(\lambda , \mu) = (1.0 , 1.0)$. 
   }
   \label{fig;alpha-falpha-1.0-1.0}
\end{figure}

\begin{figure}
\includegraphics[scale=0.6,angle=-90]{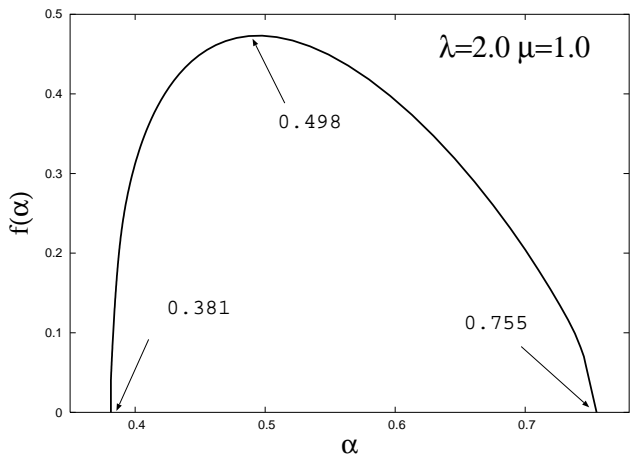}
   \caption{ $\alpha$-$f(\alpha)$ curves of the spectrum 
    for  $(\lambda , \mu) = (2.0 , 1.0)$.    }
   \label{fig;alpha-falpha-2.0-1.0}
\end{figure}

We apply the method above to the spectra of our model. 
In {\bf Fig.\ref{fig;alpha-falpha-1.0-1.0}}, we show $\alpha$-$f(\alpha)$ 
curve for the spectrum at $(\lambda,\mu)=(1.0,1.0)$. 
The estimated values of $\alpha_{\rm min}$, $\alpha_{\rm max}$ 
and $\alpha_0$ are $0.421$, $0.547$ and $0.495$ respectively. 
These values coincide with the corresponding values of the Harper 
model \cite{hiramoto}. We also investigated other points 
on lines $\lambda=2,\mu=1, \lambda=2\mu$ in {\bf Fig.\ref{fig;phase_diagram}}. 
Except for $(\lambda,\mu)=(2.0,1.0)$, it turns out that 
the estimated values of $\alpha_{\rm min}$, $\alpha_{\rm max}$ 
and $\alpha_0$ are the same as those of the Harper model. 
This implies  that the universality class for these lines 
is the same as the Harper model. This is 
consistent
 with the scaling of the total band widths (\ref{eq;totalband}) 
for $\lambda =2,\mu<1$. 
On the other hand, $\alpha$-$f(\alpha)$ 
curve for the spectrum at $(\lambda,\mu)=(2.0,1.0)$ has a different shape 
as shown in  {\bf Fig.\ref{fig;alpha-falpha-2.0-1.0}}. 
We see that $\alpha_{\rm min}=0.381$, $\alpha_{\rm max}=0.755$ and 
$\alpha_0=0.498$. Thus the universality of this point is different from 
the Harper model.

\subsection{Wavefunctions}

 We investigate  
scaling properties of the wavefunctions by the multifractal analysis. 
We concentrate on the eigenfunctions at 
   the centers of the spectra.
This enables us to confirm the phase diagram 
{\bf Fig.\ref{fig;phase_diagram}}.

\subsubsection{Multifractal Analysis of Wavefunctions} 
We apply the general formulation to characterize the wavefunctions. Take squares modulus of 
the wavefunctions to be variables $p_i$,  while take   $l_i$ 
to be constants 
\begin{equation}
l_i=l=\frac{1}{F_n} \sim  \frac{1}{\tau^n},\hspace{5mm} 
\epsilon = - \frac{1}{n} \ln l \sim  \ln \tau. 
\end{equation}
Thus $\varepsilon$ is a constant in this case. 
From (\ref{kohmoto:eq3.3}) and (\ref{kohmoto:eq3.4}), one has  
\begin{eqnarray}
G(q,\beta) = -\beta\varepsilon +G(q,0).
\label{kohmoto:eq4.2}
\end{eqnarray}
Using (\ref{kohmoto:eq3.11}) and ({\ref{kohmoto:eq3.12}), 
we obtain the generalized entropy 
\begin{eqnarray}
Q(\varepsilon,\alpha) = G(q,0)-q\varepsilon \vev{\alpha}, \hspace{5mm} 
\vev{\alpha} = -\frac{1}{\varepsilon} \frac{\partial G(q,0)}{\partial q} 
\label{kohmoto:eq4.3}. 
\end{eqnarray}
Since $Q(\varepsilon,\alpha)$ is nonzero only for $\vev{\alpha}$, 
we write it as $S'(\alpha)$. 
From (\ref{kohmoto:eq3.21}), we have    
\begin{eqnarray}
f(\alpha) = \frac{S'(\alpha)}{\varepsilon}.
\end{eqnarray}
We calculate  $f(\alpha)$ for finite Fibonacci index $n$ 
by this formula and extrapolate them to $n \rightarrow \infty$. 
Actually only a part of $f(\alpha)$ is required to 
distinguish localized, extended and critical states.  
For a localized state, $f(\alpha)$ has a point support 
and  takes  nonzero value only at $\alpha_{\rm min} =0$ and 
 $\alpha_{\rm max}=\infty$ and $f(\alpha_{\rm min})=0$, 
$f(\alpha_{\rm max})=1$. 
For an extended state, it has 
$\alpha_{\rm min}=\alpha_{max}=1$ and $f(\alpha_{\rm min })
=f(\alpha_{\rm max})=1$. 
For a critical state, $f(\alpha)$ may 
have a finite interval $[\alpha_{\rm min},\alpha_{max}]$ as 
a support and $f(\alpha)$ takes  various values.  
We shall use this method to distinguish  states near critical points.

\begin{figure}
\includegraphics[scale=0.3,angle=-90]{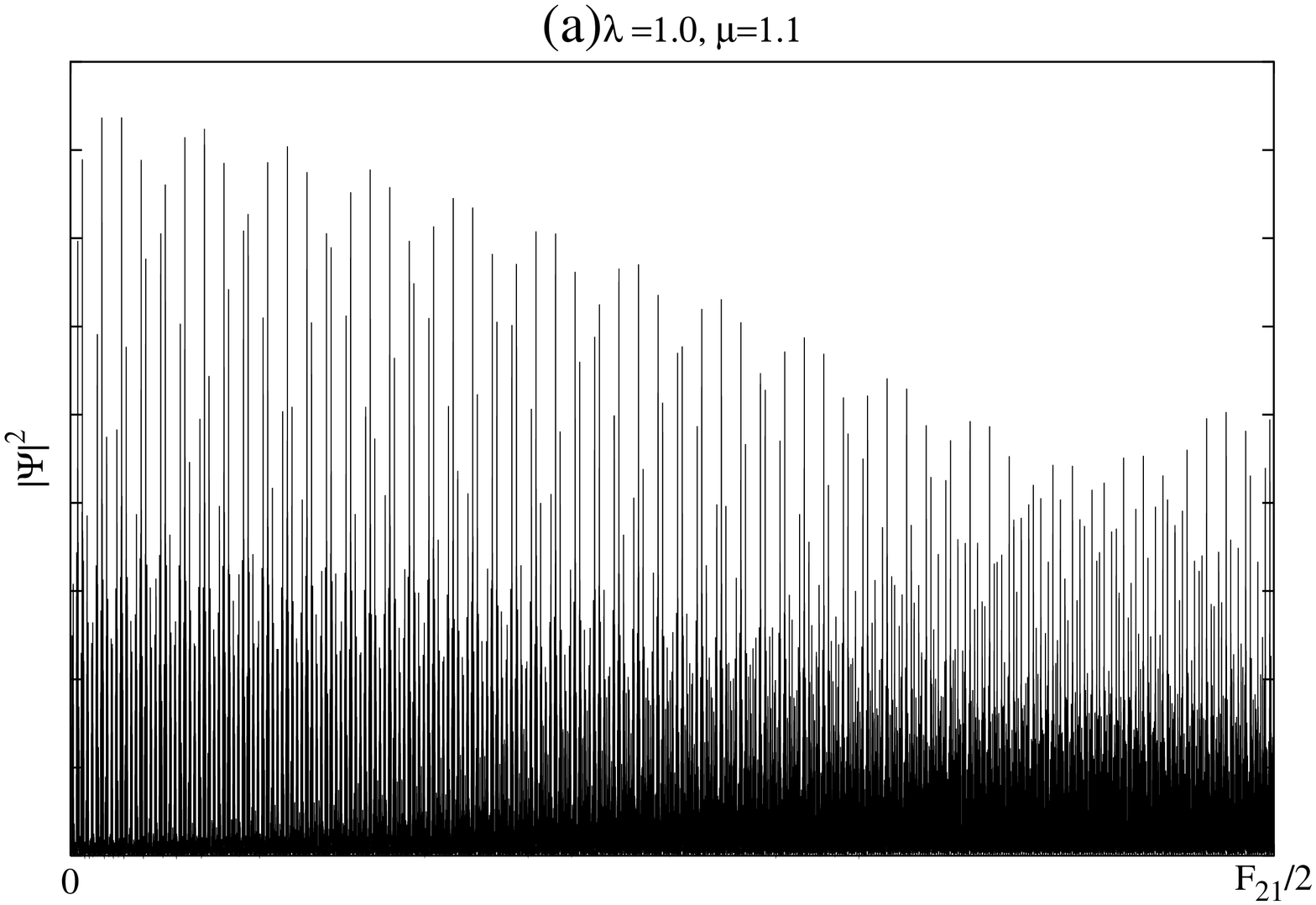}
\includegraphics[scale=0.3,angle=-90]{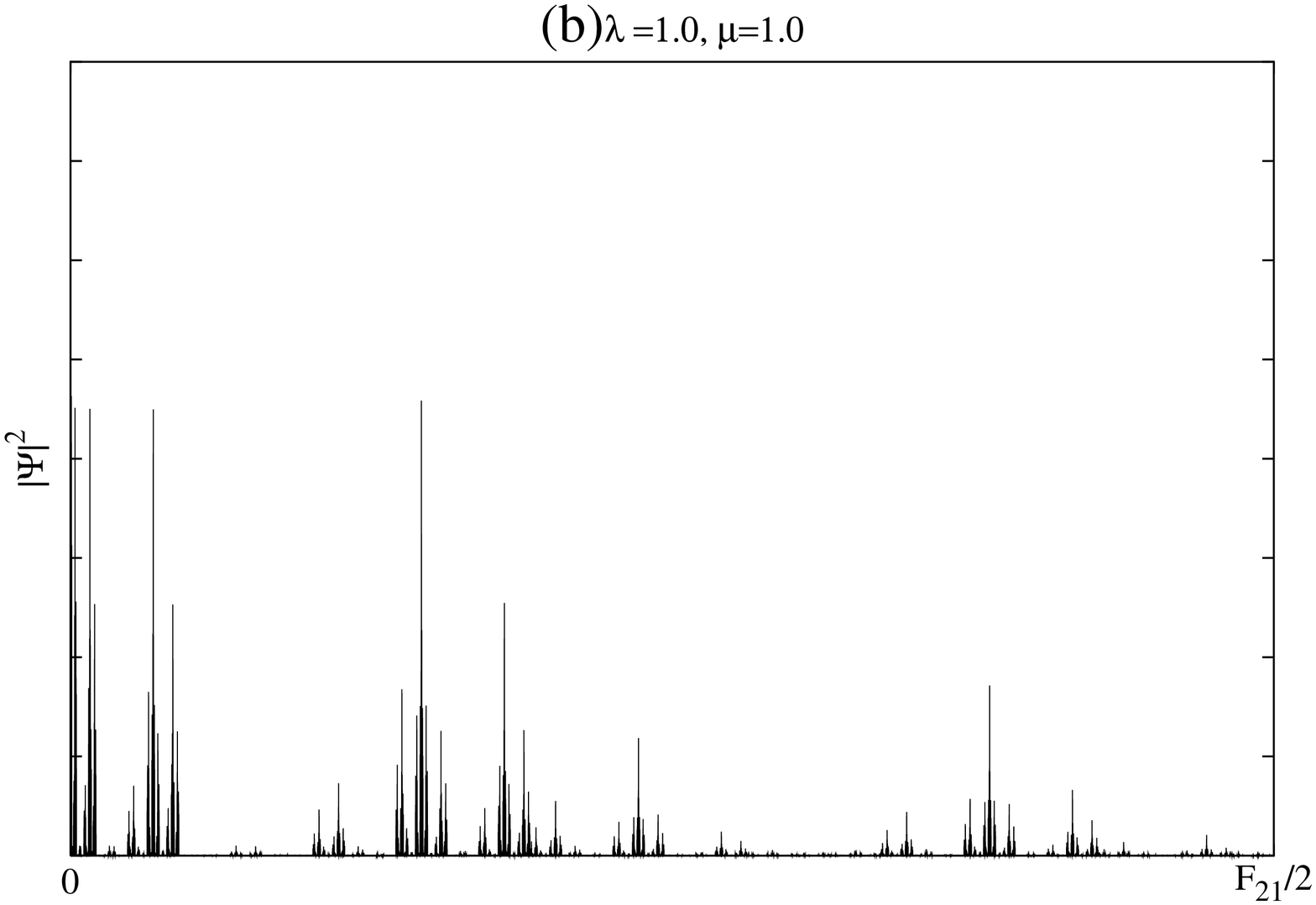}
\includegraphics[scale=0.3,angle=-90]{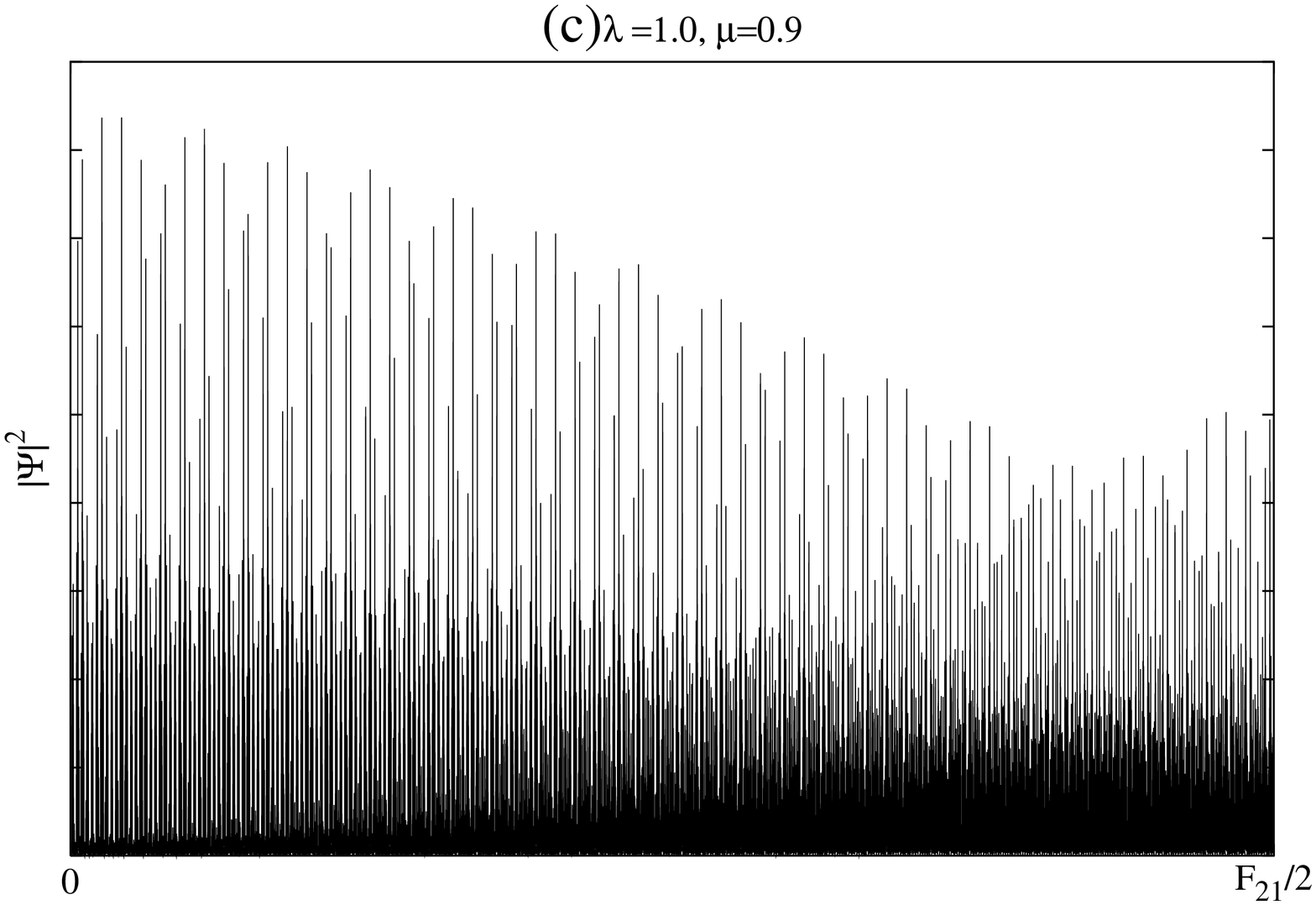}
   \caption{ Plots of 
  the wavefunctions at the centers of the spectra
for (a) $(\lambda , \mu) = (1.0 , 1.1)$ , 
(b) $(1.0,1.0)$ and
(c) $(1.0,0.9)$. 
 Here $q = 17711$.
   }
   \label{fig;wav-1.0-1.0}
\end{figure}

\begin{figure}
\includegraphics[scale=0.3,angle=-90]{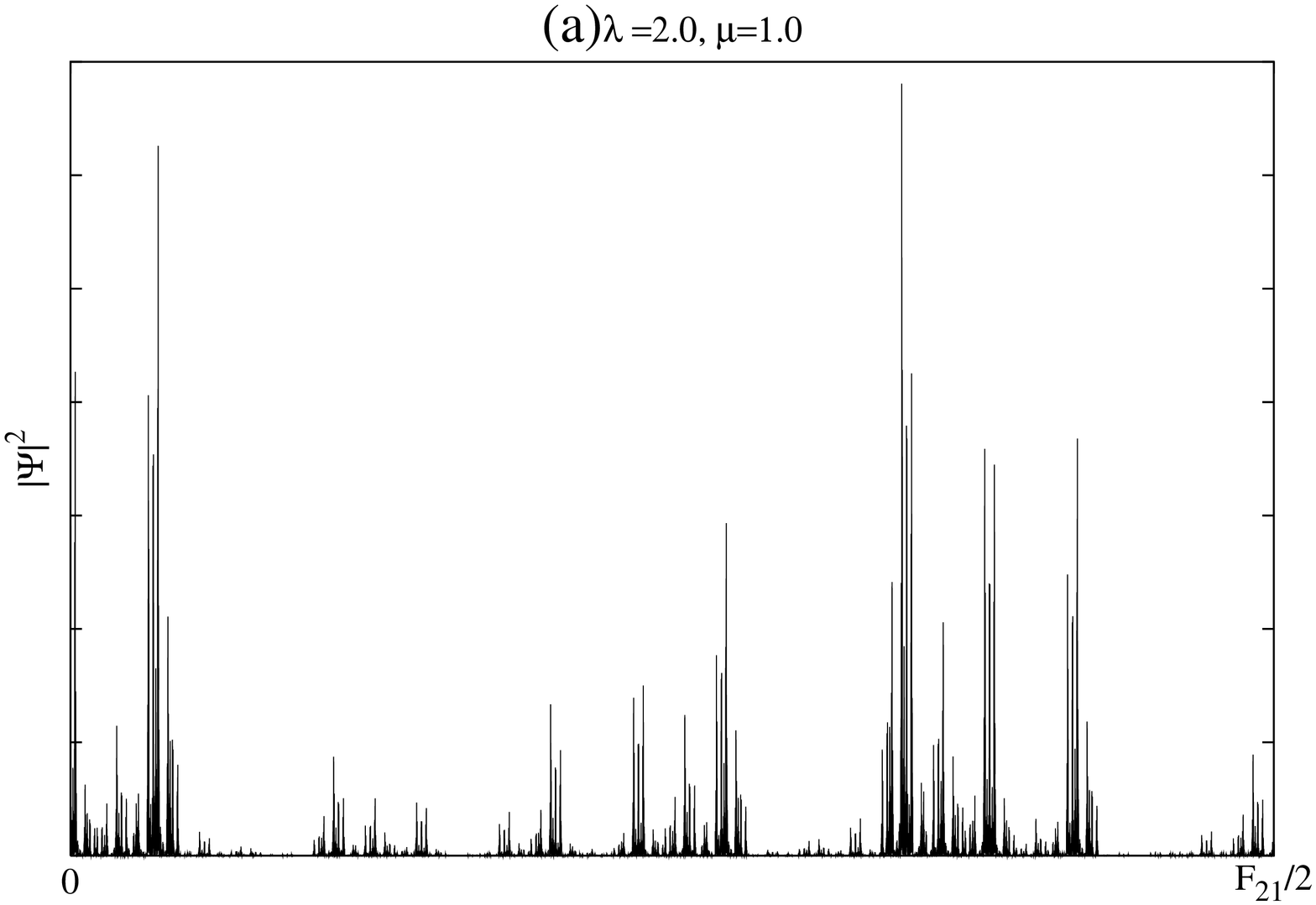}
\includegraphics[scale=0.3,angle=-90]{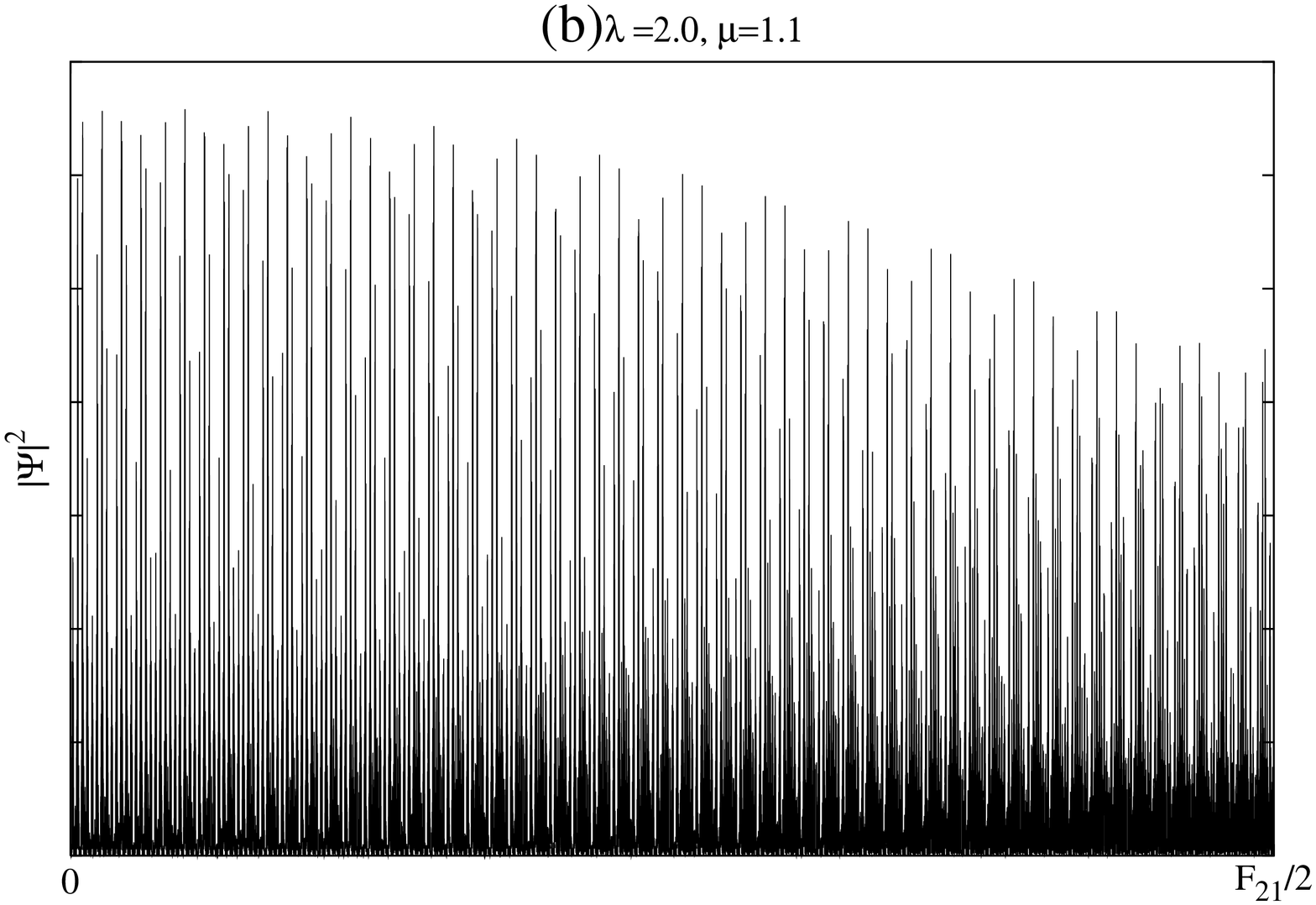}
\includegraphics[scale=0.3,angle=-90]{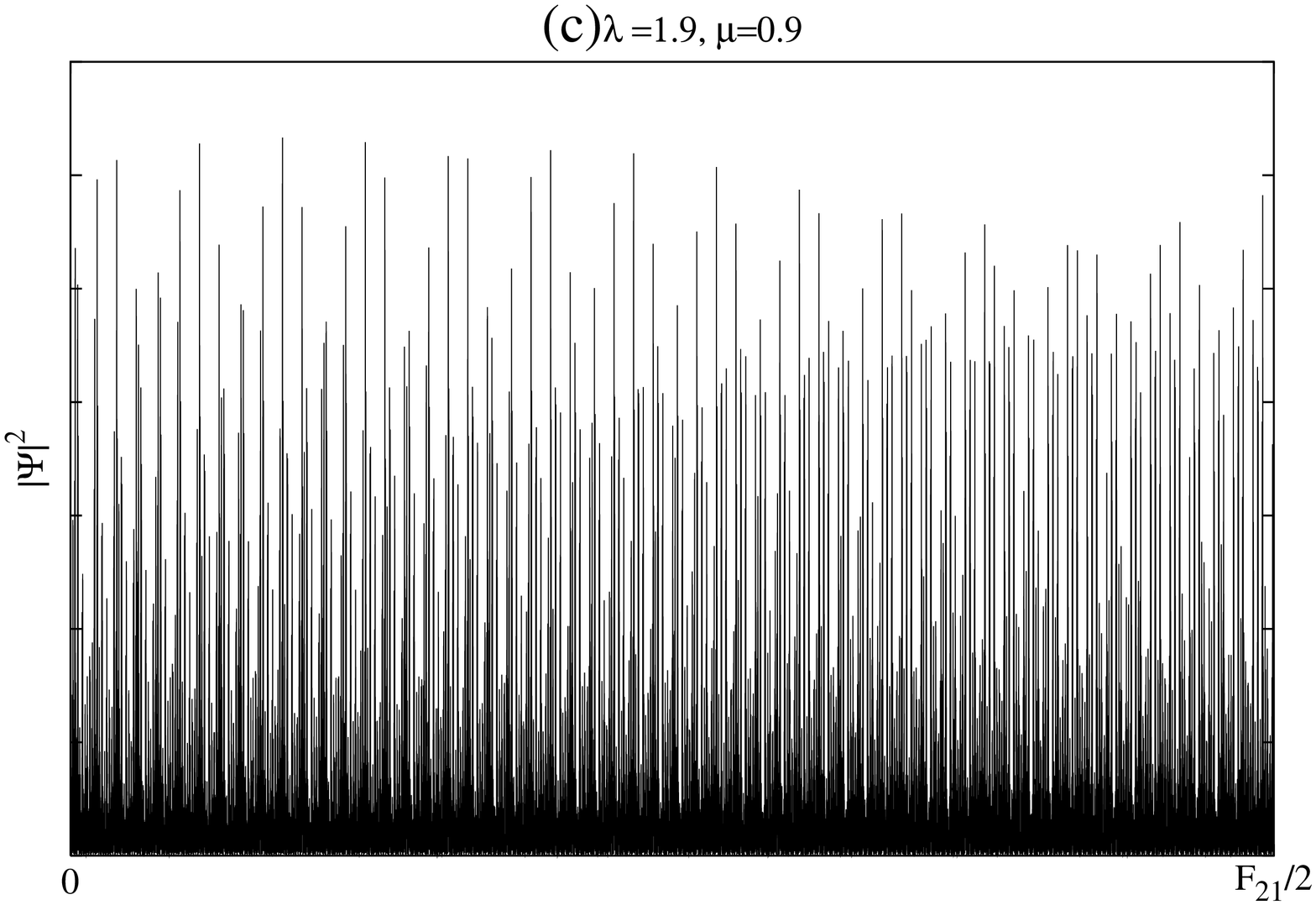}
\includegraphics[scale=0.3,angle=-90]{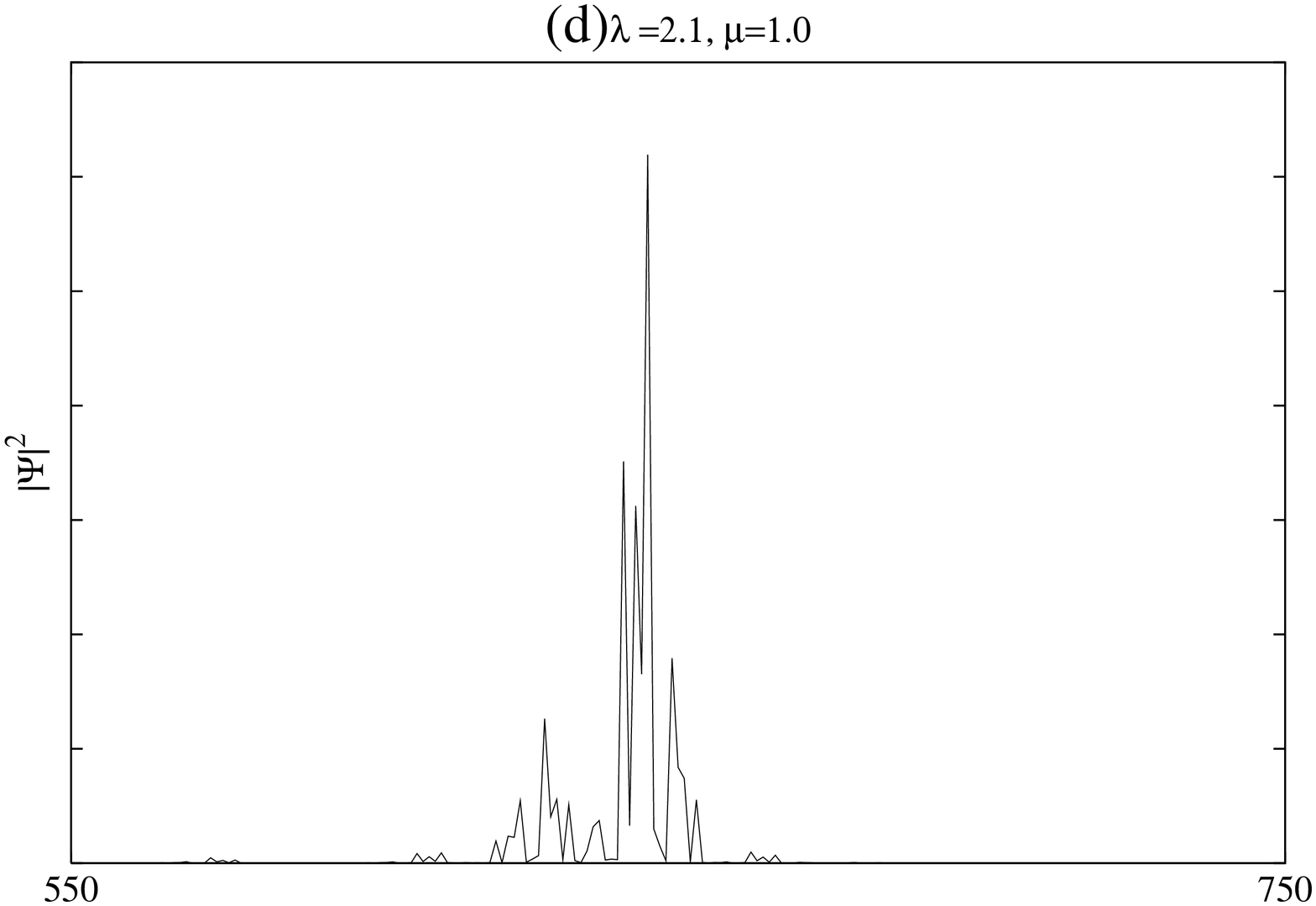}
   \caption{ Plots of  the wavefunctions at the centers of the spectra
    for (a) $(\lambda , \mu) = (2.0 , 1.0)$ , 
(b) $(2.0,1.1)$, 
(c) $(1.9,0.9)$
and (d) $(2.1,1.0)$. 
 Here $q = 17711$ for (a), (b) and (c), and  $q=4181$ for (d).
   }
   \label{fig;wav-2.0-1.0}
\end{figure}

\subsubsection{Numerical Results}
We numerically obtain the wavefunctions at the centers 
of the spectra for odd $q$ on the $\lambda=2$, $\mu=1$ and $\lambda=2\mu$ 
lines in the phase diagram {\bf Fig.\ref{fig;phase_diagram}}. 
For $\mu=1$, we investigate $q=F_n$ with $n=3\ell$ as well as 
$3\ell+1$. Although the dispersion relations are 
flat when $n=3\ell+1$, we find that 
the wavefunctions still show a characteristic 
behavior of a critical state.

In {\bf Fig.\ref{fig;wav-1.0-1.0}}, 
the square  moduli of  the wavefunctions at 
the centers of spectra for $(\lambda,\mu)=(1.0,1.1) , 
(1.0,1.0)$  and  $(1.0,0.9)$ are displayed for $n=21$ and $F_n=17711$. 
From these figures, we see that 
the wavefunctions  are extended for $(1.0,1.1)$
and for $(1.0,0.9)$, and critical for $(1.0,1.0)$
which is in accord with the phase diagram {\bf Fig.\ref{fig;phase_diagram}}. 
In {\bf Fig.\ref{fig;wav-2.0-1.0}}, the square  modulus
 of the wavefunctions at 
the band center for $(\lambda,\mu)=(2.0,1.0) , 
(2.0,1.1),(1.9,0.9)$ and $(2.1,1.0)$ i.e. in the vicinity of  the 
bicritical point of {\bf Fig.\ref{fig;phase_diagram}} 
are displayed  for $n=21$ and $F_n=17711$ 
($n=18$ and $F_n=4181$ for $(2.1,1.0)$).
It is rather clear that 
the wavefunction is extended for $(2.0,1.1)$ and $(1.9,0.9)$, 
localized for $(1.0,0.9)$ and critical for $(2.0,1.0)$. 
To draw convincing conclusions, however,
it is necessary to study  the scaling properties by multifractal analysis.

We plot $\alpha_{\rm min}$ for $(\lambda,\mu)=(1.0,1.1), 
(1.0,1.0)$  and  $(1.0,0.9)$   in   {\bf Fig.\ref{fig;alpha-1.0-1.0}}. 
 For $(\lambda,\mu)=(1.0,0.9)$ and  $(1.0,1.1)$, 
it is clearly seen that $\alpha_{\rm min}$ extrapolates to $1$ for 
$n \rightarrow \infty$. On the other hand, $\alpha_{\rm min}$  
 is extrapolated to $0.358$ for $(\lambda,\mu)=(1.0,1.0)$.
This value of $\alpha_{\rm min}$ is actually the same as 
the one found in the Harper model \cite{hiramoto} within 
statistical error.   
As shown in {\bf Fig.\ref{fig;falpha-1.0-1.0}} 
$f(\alpha_{\rm min})$ extrapolates to $1$ for $(\lambda,\mu)=(1.0,1.1)$
and $(1.0,0.9) $, and $0$ for $(1.0,1.0)$.  
The behaviors of $\alpha_{\rm min}$ and $f(\alpha_{\rm min})$  in 
{\bf Fig.\ref{fig;alpha-1.0-1.0} -\ref{fig;falpha-1.0-1.0} }
indicate that the state is extended 
for $(\lambda,\mu)=(1.0,1.1)$ and $(1.0,0.9) $, and critical for $(1.0,1.0)$.   
This confirms a part of 
the phase diagram in {\bf Fig.\ref{fig;phase_diagram}}, 
especially the metal-metal transitions at $\mu=1.0$.

\begin{figure}
\includegraphics[scale=0.3,angle=-90]{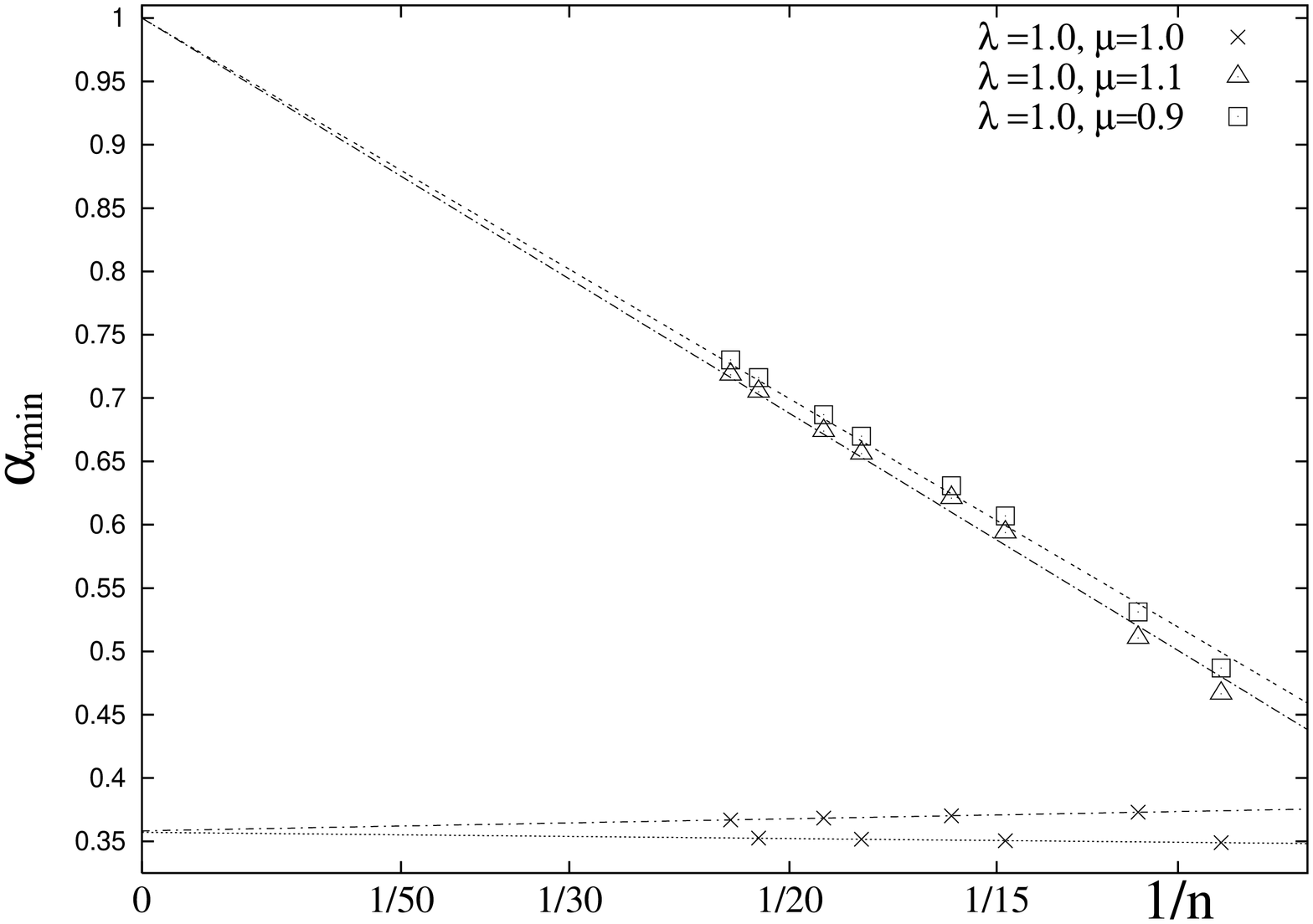}
   \caption{ Plots of $\alpha_{\rm min}$ vs. $\frac{1}{n}$ 
near $(\lambda,\mu)=(1.0,1.0)$ with  
$n=12,13,15,16,17,18,19,21$ and $22$.  }
   \label{fig;alpha-1.0-1.0}
\end{figure}

\begin{figure}
\includegraphics[scale=0.3,angle=-90]{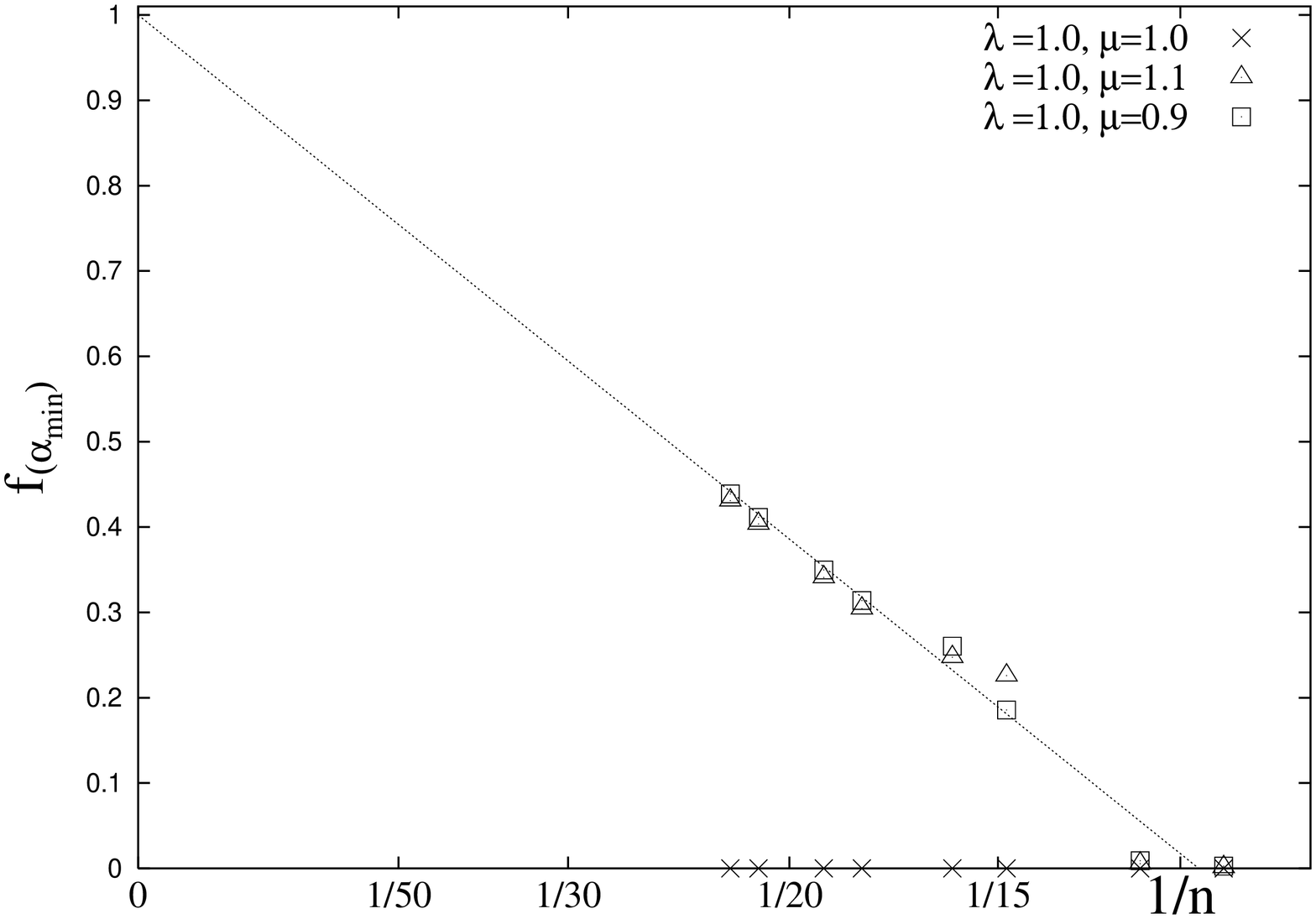}
   \caption{ Plots of $f(\alpha_{\rm min})$ against $\frac{1}{n}$ near
 $(\lambda,\mu)=(1.0,1.0)$.  }
   \label{fig;falpha-1.0-1.0}
\end{figure}

\begin{figure}
\includegraphics[scale=0.3,angle=-90]{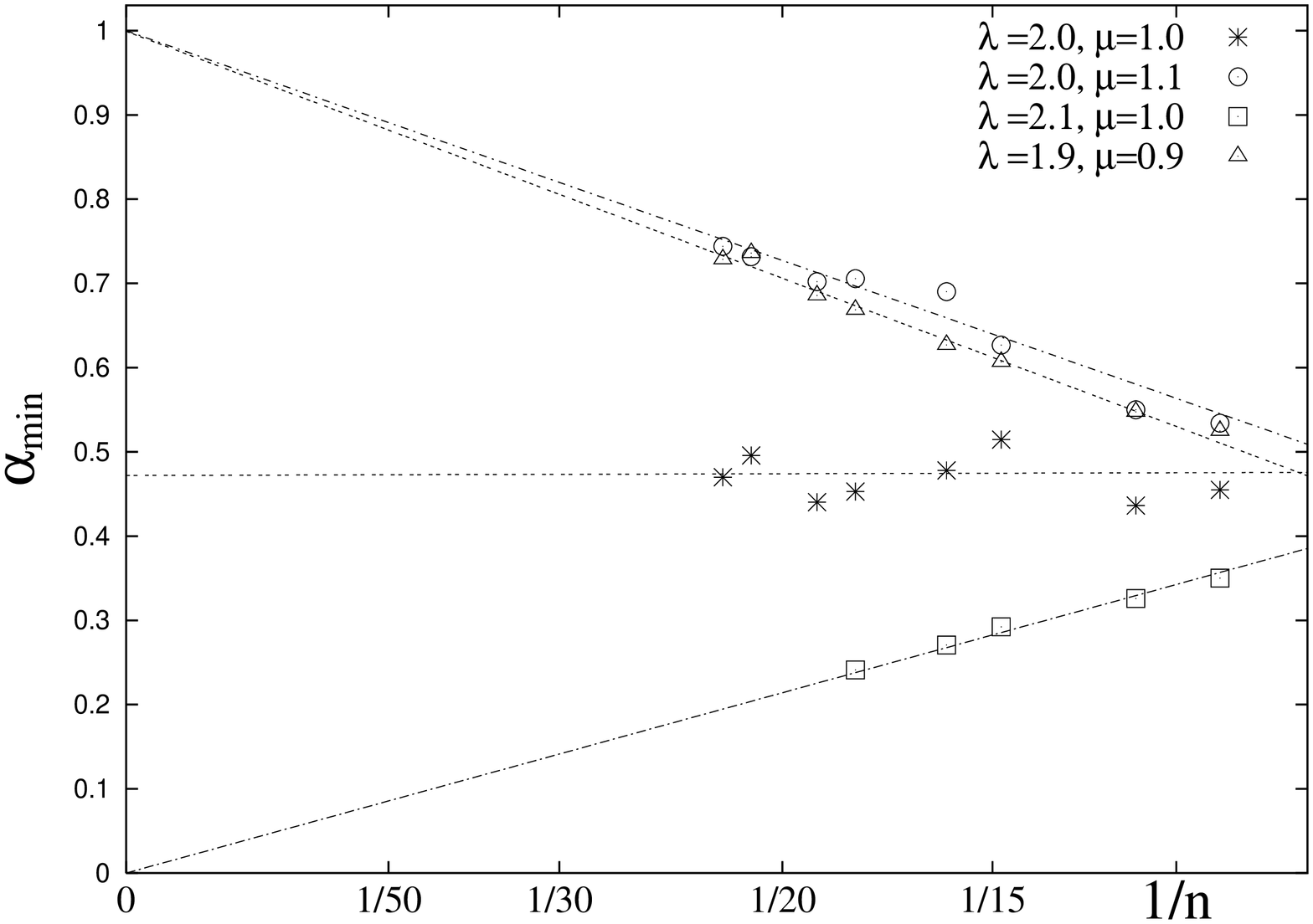}
   \caption{ Plots of $\alpha_{\rm min}$ against $\frac{1}{n}$ 
near $(\lambda,\mu)=(2.0,1.0)$.  }
   \label{fig;alpha-2.0-1.0}
\end{figure}

\begin{figure}
 \includegraphics[scale=0.3,angle=-90]{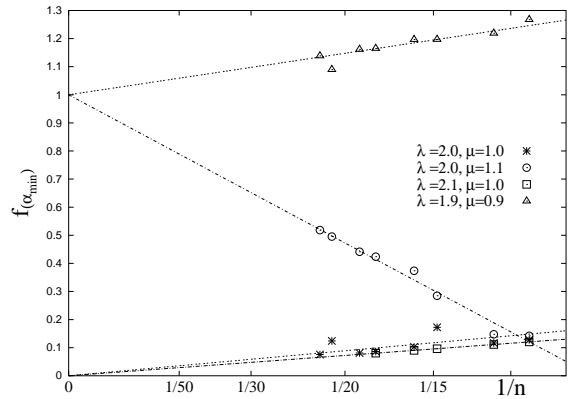}
   \caption{ Plots of $f(\alpha_{\rm min})$ vs. $\frac{1}{n}$ 
near $(\lambda,\mu)=(2.0,1.0)$.  }
   \label{fig;falpha-2.0-1.0}
\end{figure}

\begin{figure}
\includegraphics[scale=0.3,angle=-90]{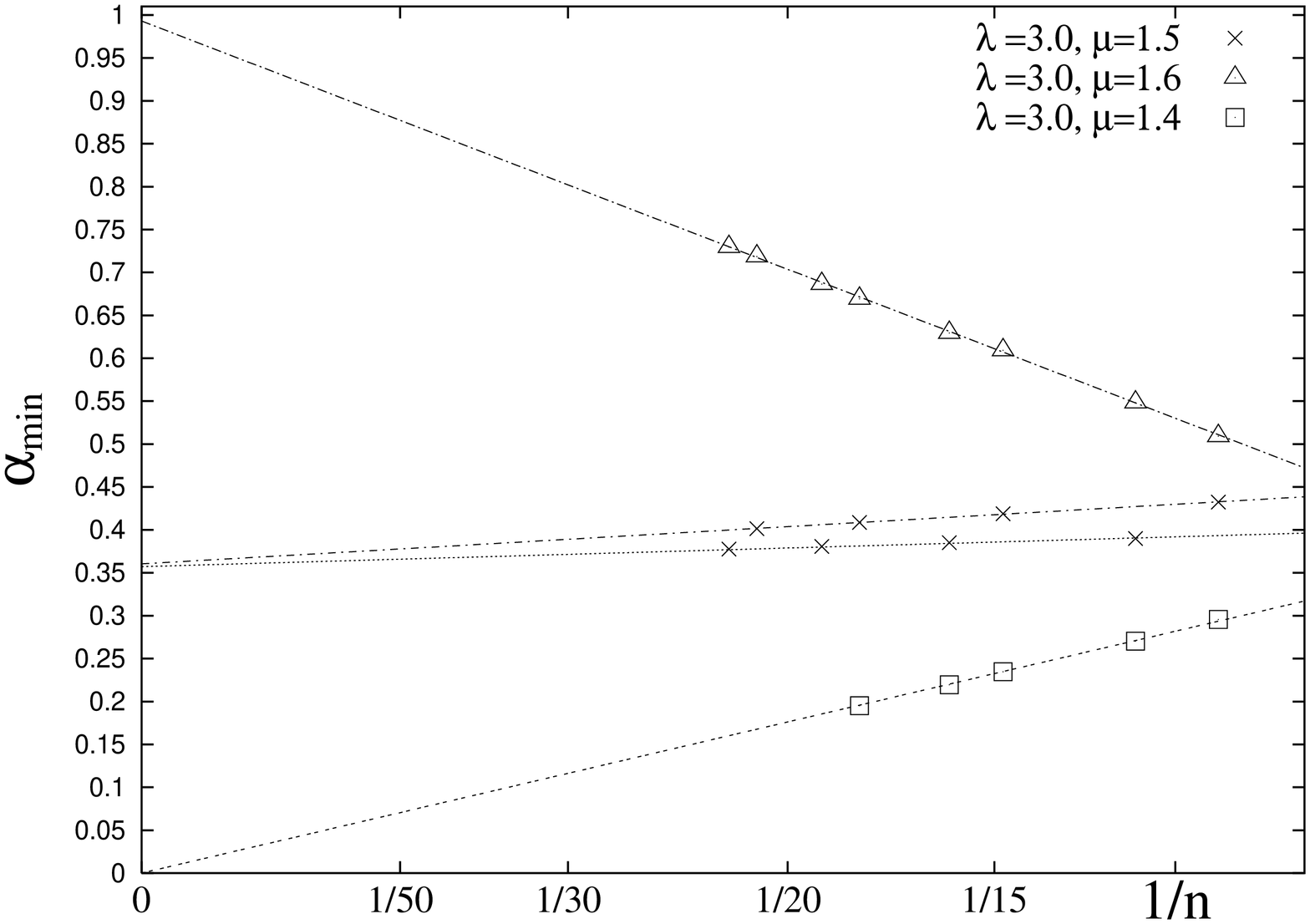}
   \caption{ Plots of $\alpha_{\rm min}$ against $\frac{1}{n}$ 
near $(\lambda,\mu)=(3.0,1.5)$.  }
   \label{fig;alpha-3.0-1.5}
\end{figure}

\begin{figure}5
 \includegraphics[scale=0.3,angle=-90]{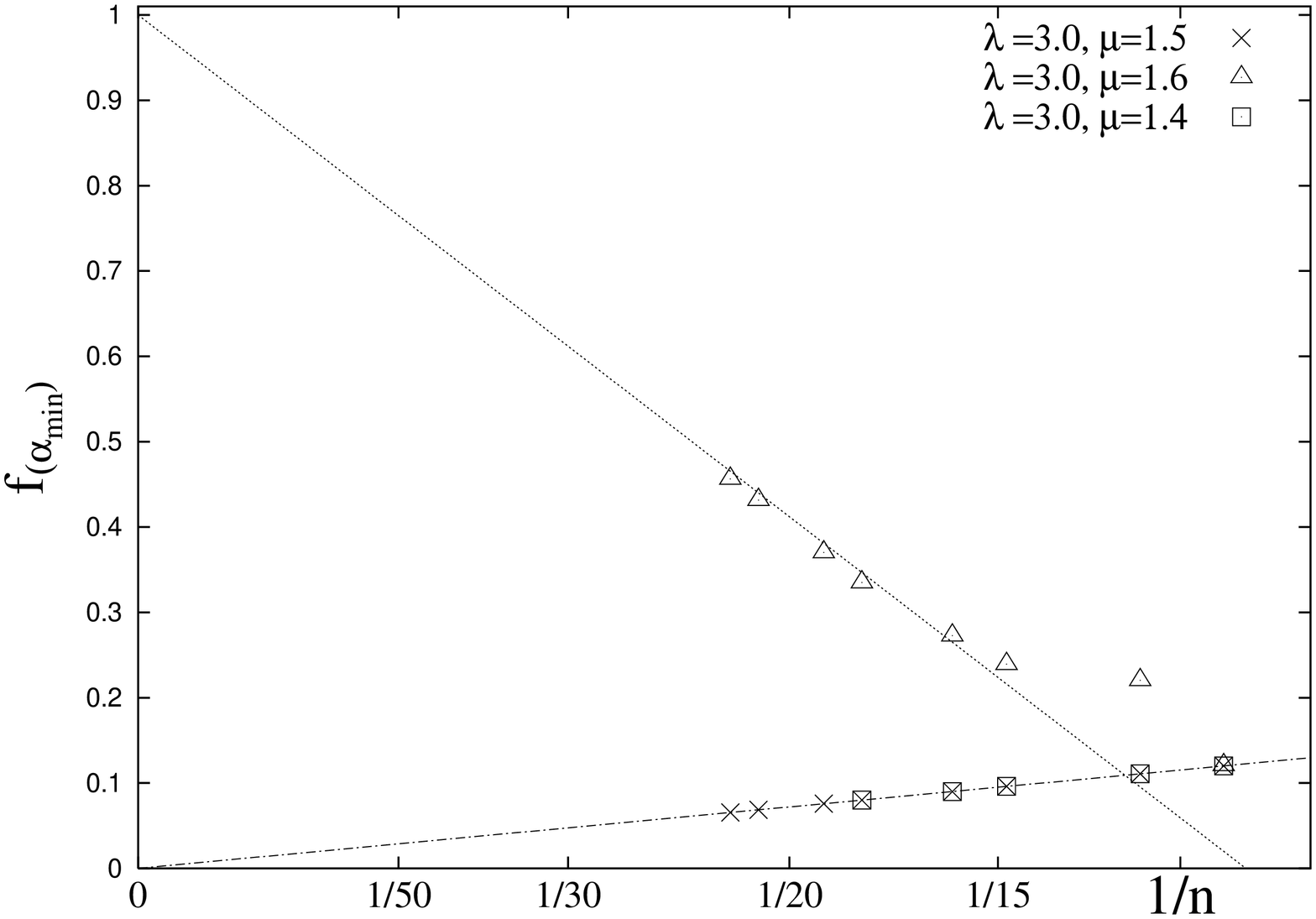}
   \caption{ Plots of $f(\alpha_{\rm min})$ against $\frac{1}{n}$ 
near $(\lambda,\mu)=(3.0,1.5)$.  }
   \label{fig;falpha-3.0-1.5}
\end{figure}

In {\bf Fig. \ref{fig;alpha-2.0-1.0}},   $\alpha_{\rm min}$'s are shown for the
states near the bicritical point $(\lambda,\mu)=(2.0,1.0), 
(2.0,1.1),(1.9,0.9)$ and $(2.1,1.0)$.  Also 
{\bf Fig. \ref{fig;falpha-2.0-1.0}} shows  $f(\alpha_{\rm min})$'s for 
them.
For $(2.0,1.1)$ and $(1.9,0.9)$, both $\alpha_{\rm min}$ and 
$f(\alpha_{\rm min})$ are extrapolated to $1$, telling that 
the state is extended. For $(2.1,1.0)$, both $\alpha_{\rm min}$ and 
$f(\alpha_{\rm min})$ are extrapolated to $0$, which means that 
the state is localized. At $(2.0,1.0)$,  the convergence of $\alpha_{\rm min}$ seems slow 
but the plots show a tendency to converge to a finite value 
near $0.47$. Similarly $f(\alpha_{\rm min})$  converges 
to zero. Thus we conclude that the states are critical at 
$(2.0,1.0)$  and it is the bicritical point of 
  metal-insulator and metal-metal transitions. See
{\bf Fig.{\ref{fig;phase_diagram}}}. 

Next {\bf Fig.{\ref{fig;alpha-3.0-1.5}}}  and 
{\bf Fig.{\ref{fig;falpha-3.0-1.5}}} show 
the scaling of $\alpha_{\rm min}$ and $f(\alpha_{\rm min})$ respectively 
near $(\lambda,\mu)=(3.0,1.5)$.  The extrapolated values of 
$\alpha_{\rm min}$ and $f(\alpha_{\rm min})$ are consistent with 
the phases diagram {\bf Fig.{\ref{fig;phase_diagram}}}. 
We investigate $\alpha_{\rm min}$ and $f(\alpha_{\rm min})$ 
for other points on the critical lines and the results are consistent with the $(\lambda,\mu)$-  phase diagram
 {\bf Fig.{\ref{fig;phase_diagram}}}

\begin{figure}
\includegraphics[scale=0.6,angle=-90]{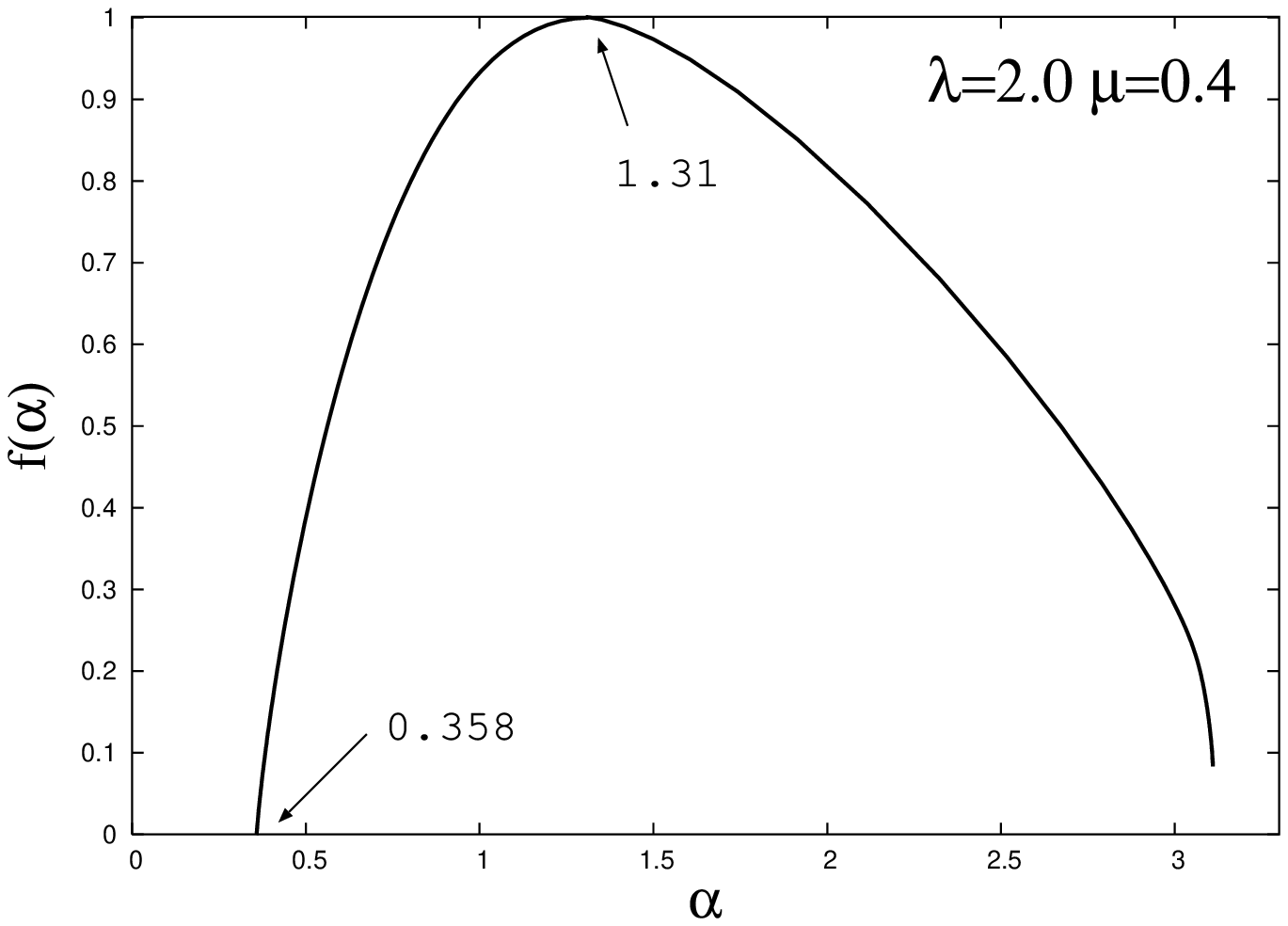}
   \caption{ $\alpha$-$f(\alpha)$ curve for  $(\lambda,\mu)=(2.0,0.4)$.  }
   \label{fig;alpha-falpha-2.0-0.4}
\end{figure}

Let us next turn to whole $\alpha$-$f(\alpha)$ curve. 
 In {\bf Fig.\ref{fig;alpha-falpha-2.0-0.4}} 
 the $\alpha$-$f(\alpha)$ curve at  
$(\lambda,\mu)=(2.0,0.4)$ is shown. Within statistical error, $\alpha_{\rm min}$ is 
$0.358$.
This value of $\alpha_{\rm min}$ holds  
on the critical lines except 
in the vicinity of the bicritical point, 
where $\alpha_{\rm min}$ slightly changes about $0.05$. 
In {\bf Fig.\ref{fig;alpha-falpha-2.0-0.4}}, 
the value of $\alpha_0$  which gives the maximum 
of $f(\alpha)$ is observed to be $1.31$  which is also 
the same as the Harper model \cite{hiramoto}. 
It is the same 
for  the other points on the critical lines $\lambda=2.0$ 
and $\lambda=2\mu$ within 
statistical error, except in the vicinity of the bicritical point 
where $\alpha_0$ slightly changes within $0.05$.  
We may interpret these slight changes of $\alpha_{\rm min}$ and $\alpha_0$  
in the vicinity of the bicritical point as  an effect of  
slow convergence there (see below). 
For $f(\alpha)$ for $\alpha > \alpha_0$, 
the convergence of $f(\alpha)$ is not good, especially  
for $\alpha_{\rm max}$.

For the critical line $\mu=1$, $\alpha_0$ splits for $n=3 \ell$  
and $n=3 \ell +1 $.  In {\bf Fig.\ref{fig;alpha0-1.0-1.0}},  
plots of $\alpha_0$ and $\alpha_{\rm min}$ for 
$(\lambda,\mu) =(1.0,1.0)$ are shown.   
It is seen that $\alpha_0$ goes to different 
values for  $n=3\ell$  and $n=3\ell+1$. Similar 
behavior of $\alpha_0$ is observed for other points 
on the $\mu=1.0$ critical line.  
This implies that the universality class is different 
for these two series. 
Also the $\alpha$-$f(\alpha)$ curves for $n=3\ell$ and $ 3\ell+1$ are 
shown in {\bf Fig.\ref{fig;alpha-falpha-1.0-1.0-wav}}.  
 The estimated values of $\alpha_{\rm min}$ and $\alpha_0$ 
for $n=3\ell$ and $3\ell+1$ with  $\mu=1$ are shown in {\bf Table \ref{table;wav-alpha}} .  

For the bicritical point $(\lambda,\mu)=(2.0,1.0)$, 
the split of $3\ell$ and $3\ell+1$ is not so obvious in 
the numerical data.  For example, 
{\bf Fig. \ref{fig;alpha-2.0-1.0}} shows 
$\alpha_{\rm min}$ for $n=12,13,\cdots$ and they don't  
clearly split into two series. Also the convergence 
is not as good as those for other points. 
The numeric for $(\lambda,\mu)=(2.0,1.0)$ in 
{\bf Table \ref{table;wav-alpha}} is obtained based 
on both series.

\begin{figure}
\includegraphics[scale=0.6]{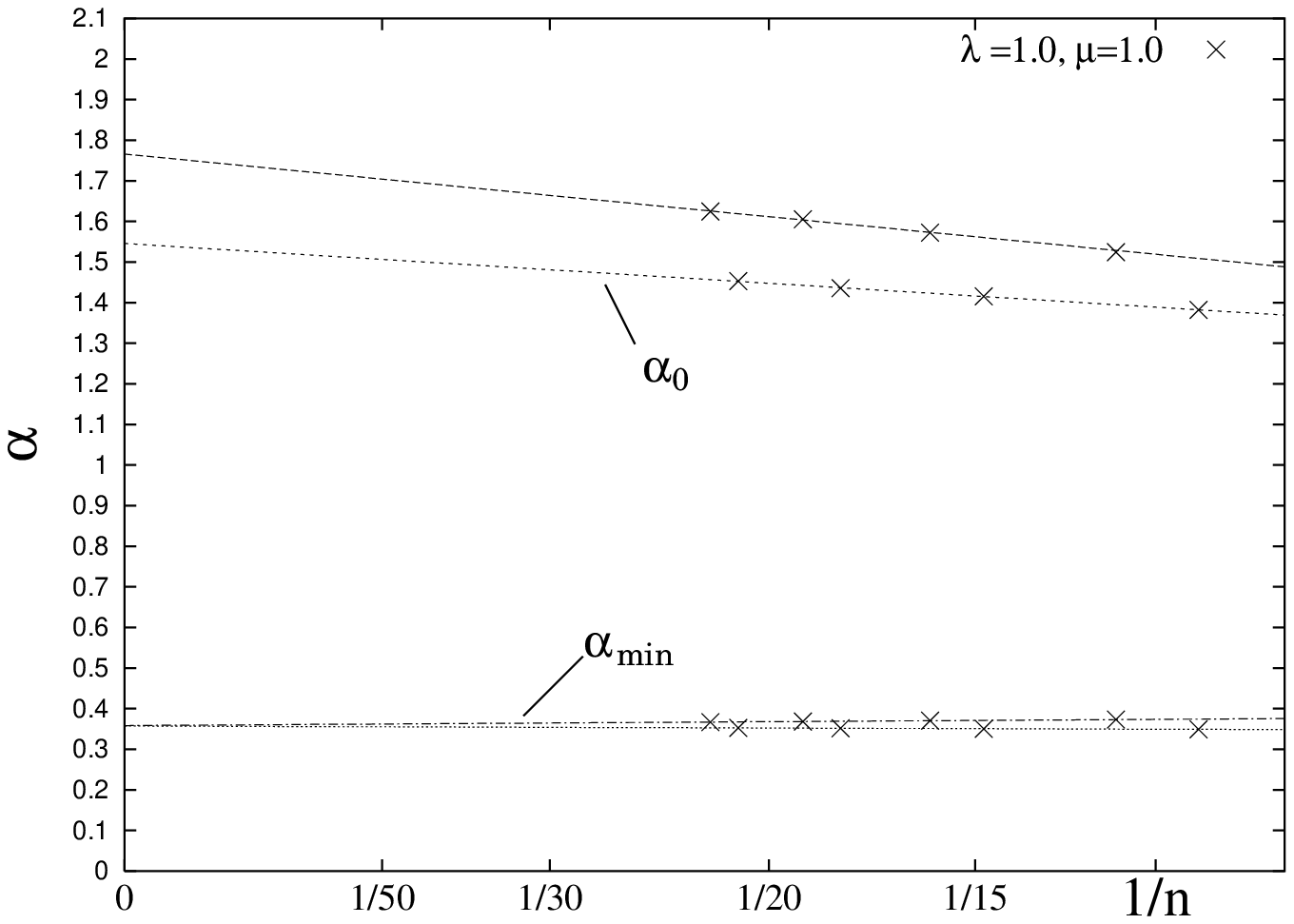}
   \caption{ The upper two series are plots of $\alpha_0$ and 
the lower two series are plots of $\alpha_{\rm min}$ 
$(\lambda,\mu)=(1.0,1.0)$.  }
   \label{fig;alpha0-1.0-1.0}
\end{figure}

\begin{figure}
\includegraphics[scale=0.5,angle=-90]{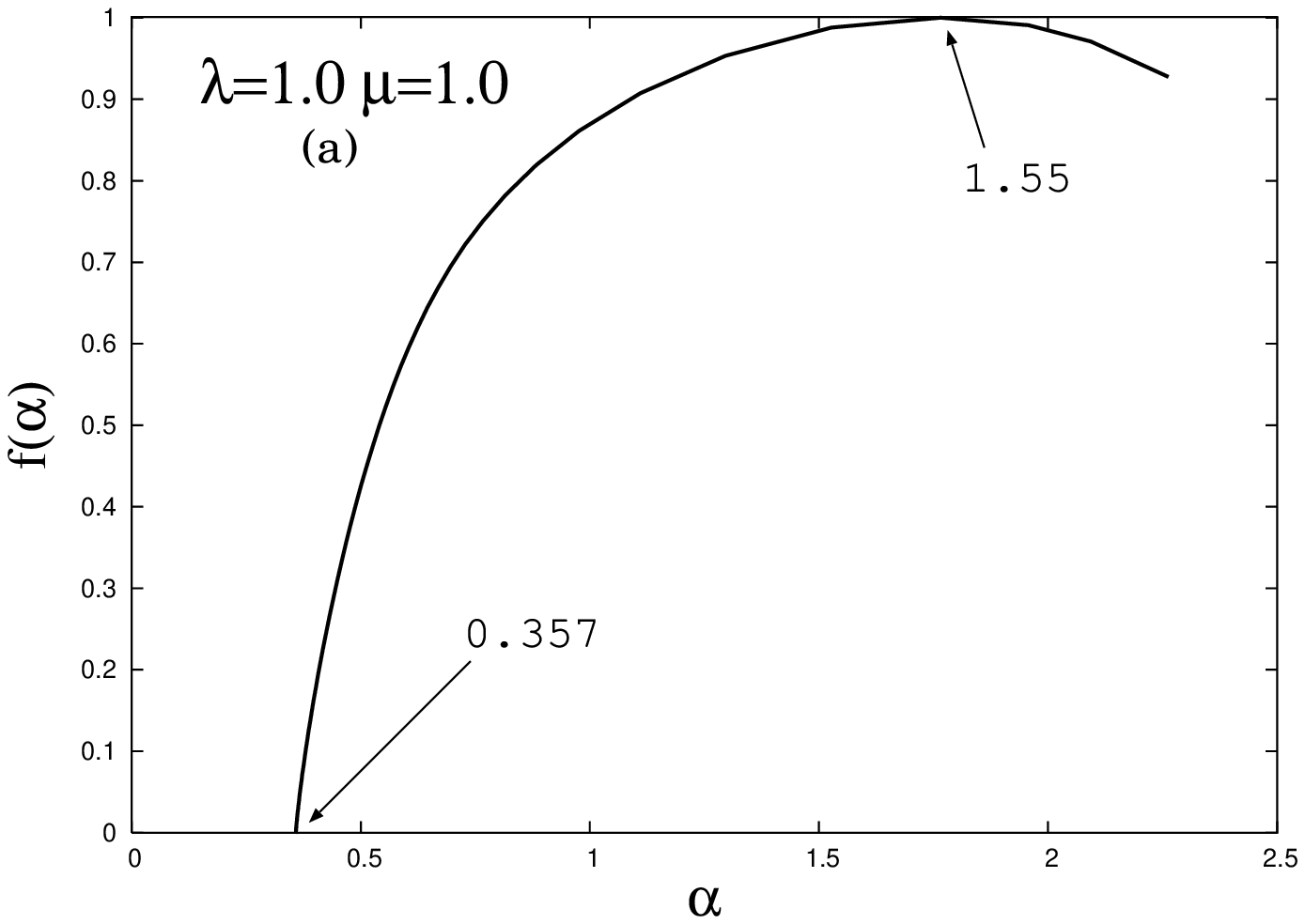}
\includegraphics[scale=0.5,angle=-90]{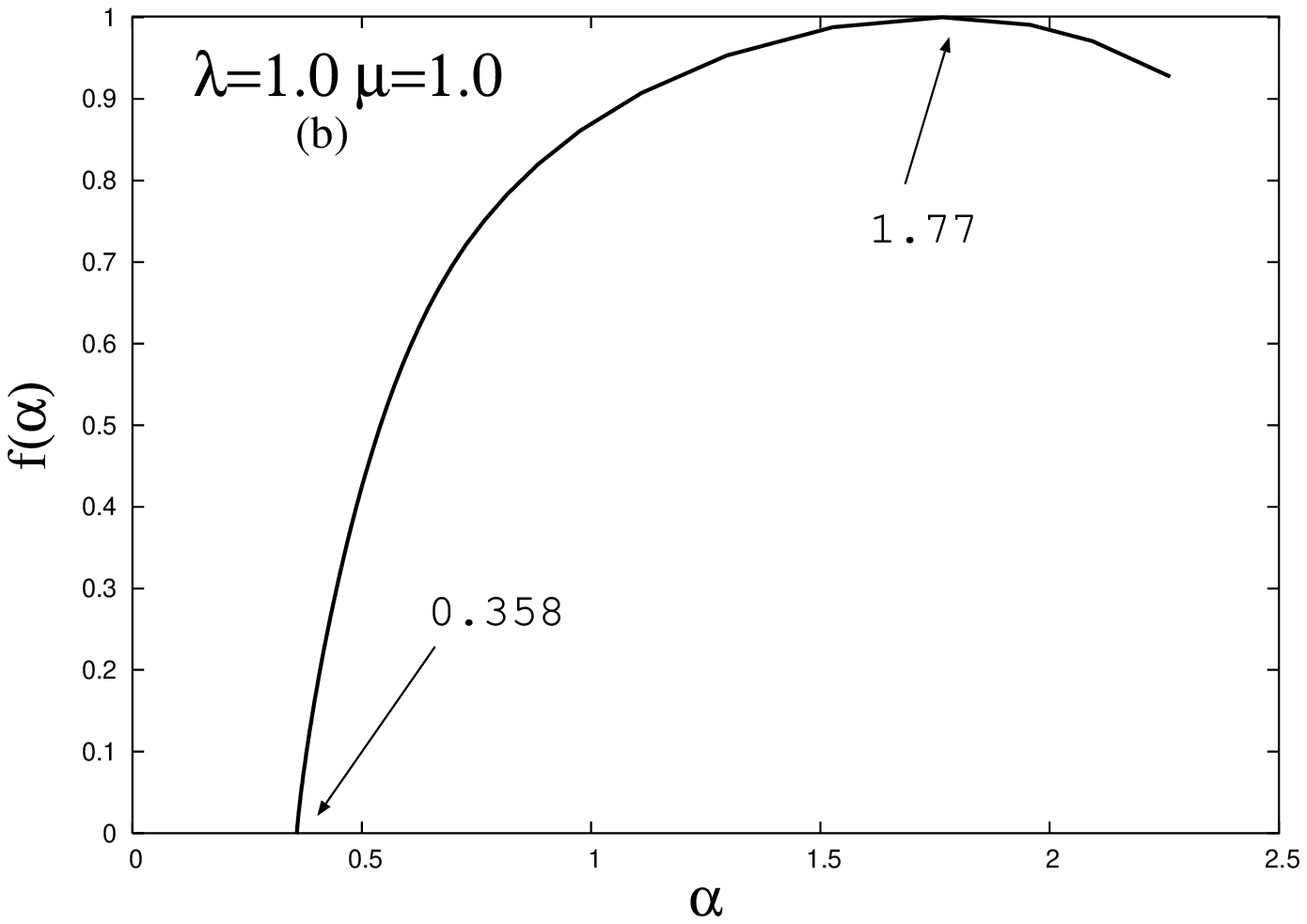}
   \caption{ $\alpha$-$f(\alpha)$ curve at  $(\lambda,\mu)=(1.0,1.0)$, 
(a) for index=12,15,18 and 21 and (b) for index=13,16,19 and 22.   }
   \label{fig;alpha-falpha-1.0-1.0-wav}
\end{figure}

\setlength{\arrayrulewidth}{0.8pt}
\begin{table}[tb]
\begin{tabular}{ccccc}     \hline
index & $\lambda$ & $\mu$ & $\alpha_{\rm min}$ & $\alpha_0 $  \\  \hline
$3\ell$ &0.0 & 1.0 & 0.358 & 1.82    \\
& 0.5 & 1.0 & 0.358 & 1.57    \\
&1.0 & 1.0 & 0.357  & 1.55     \\
&1.5 & 1.0 & 0.357  & 1.54   \\
\hline
$3\ell+1$& 0.0 & 1.0 & 0.358 & 1.65    \\
&0.5 & 1.0 & 0.358   & 1.77     \\
&1.0 & 1.0 & 0.358  & 1.77     \\
&1.5 & 1.0 & 0.358  & 1.73     \\
\hline 
&2.0&1.0& 0.47 & 1.4
\end{tabular}
\caption{
$\alpha_{\rm min}$ and $\alpha_0$ on the critical line 
$\mu=1.0$.}
\label{table;wav-alpha}
\end{table}

\section{\label{section;conclusions}Conclusions}
We study two dimensional electrons on the 
triangular lattice in a  uniform magnetic field and 
 the one dimensional quasiperiodic system obtained from it. 
We conjectured a phase diagram of the one dimensional model 
as in {\bf Fig. \ref{fig;phase_diagram}}. As a typical example, 
we investigated the incommensurate limit of the golden mean 
via  
the level statistics, namely the distributions of the band widths and the gaps, 
and scaling properties of spectra and wavefunctions on 
the conjectured critical lines. For level statistics, 
we find the characteristic behaviors similar to the ones 
previously found for other quasiperiodic models. 
We also obtain $\alpha$-$f(\alpha)$ curve for 
spectra and the wavefunctions at the centers  of the spectra.  As for the spectra, 
 $\alpha$-$f(\alpha)$ curve is the same as one in the Harper model 
 on the critical lines  except for the bicritical point. 
 For the wavefunctions, we find that 
 $\alpha_{\rm min}$ is the same as the Harper
 model except near the bicritical point. For the line $\lambda=2$, 
 $\alpha$-$f(\alpha)$ curves are the same as 
  the Harper model. 
For $\mu$=1, the Fibonacci sequence $F_n$ splits into two classes 
$n=3\ell$ and $n=3\ell+1$ 
according to the appearance of zero of the hopping terms.
The dispersion relation is flat for $n=3\ell+1$. 
Also  the $\alpha$-$f(\alpha)$ curve of  the wavefunction is 
 different for $n=3\ell$ and $n=3\ell+1$. 
At the bicritical point where the triangular lattice 
symmetry retains,  both level statistics and multifractal analysis 
show qualitively different behaviors from those of other critical points.

{\it Acknowledgement} 
K. I. would like to thank Y. Takada for collaboration at the 
initial stage of this work.  
K.I. was partially supported by 
the Grand-in-Aid for Science(B), No.1430114  of JSPS.

\end{document}